\documentclass[journal,twoside,web]{ieeecolor}
\usepackage{generic}
\usepackage{cite}
\usepackage{amsmath,amssymb,amsfonts}
\usepackage{algorithmic}
\usepackage{graphicx}
\usepackage{algorithm,algorithmic}
\usepackage{hyperref}
\hypersetup{hidelinks=true}
\usepackage{textcomp}
\def\BibTeX{{\rm B\kern-.05em{\sc i\kern-.025em b}\kern-.08em
    T\kern-.1667em\lower.7ex\hbox{E}\kern-.125emX}}
\usepackage{bm}
\usepackage{array}
\usepackage{mathtools}
\definecolor{myblue}{rgb}{0,0,0.65}
\definecolor{myred}{rgb}{0.75,0,0}
\definecolor{mygreen}{rgb}{0,0.5,0}
\definecolor{mypurple}{rgb}{0.5, 0, 0.5}

\newtheorem{theorem}{Theorem}
\newtheorem{corollary}{Corollary}
\newtheorem{lemma}{Lemma}
\newtheorem{definition}{Definition}
\newtheorem{assumption}{Assumption}
\newtheorem{remark}{Remark}

\newcommand{\LL}{\mathcal{L}}

\newcommand{\zero}
{
\mathbf{0}
}

\newcommand{\one}
{
\mathbf{1}
}

\newcommand*{\Rx}[1]
{
\mathbb{R}^{#1}
}

\newcommand*{\Cx}[1]
{
\mathbb{C}^{#1}
}

\newcommand{\Cxx}[2]
{
\mathbb{C}^{#1 \times #2}
}

\newcommand{\Tr}
{
\operatorname{Tr}
}

\newcommand{\norm}[1]
{
\left\lVert{#1}\right\rVert
}

\newcommand{\Ein}
{
{\bm{E}_{\text{in}}}
}

\newcommand{\Fin}
{
{\bm{F}_{\text{in}}}
}

\newcommand{\Gin}
{
{\bm{G}_{\text{in}}}
}

\newcommand{\Din}
{
{\bm{D}_{\text{in}}}
}

\begin{document}

\title{Frequency Response Identification of Low-Order Systems: Finite-Sample Analysis}

\author{Arya Honarpisheh, Mario Sznaier
\thanks{This work was partially supported by NSF grants CNS--2038493 and CMMI--2208182, AFOSR grant FA9550-19-1-0005 and ONR grant N00014-21-1-2431.}
\thanks{Arya Honarpisheh and Mario Sznaier are with the Robust Systems Lab, ECE Department, Northeastern University, Boston, MA 02115 USA (e-mail: honarpisheh.a@northeastern.edu, msznaier@coe.neu.edu).}}

\maketitle

\pagestyle{plain}
\thispagestyle{empty}

\begin{abstract}
This paper proposes a frequency-domain estimator for low-order systems from repeated noisy measurements. The estimator minimizes a quadratic data-fitting term regularized by the nuclear norm of a Loewner matrix, subject to a convex stability constraint enforced via a semidefinite program. We prove a finite-sample error bound at the sampled frequencies and extend it to all frequencies through rational interpolation. The bound characterizes the dependence on the number of repeated experiments, number of frequency points, system order, and noise level. Numerical experiments on SISO and MIMO systems demonstrate the low-order-promoting effect of the method and validate the predicted scaling laws. \footnote{Code available at \url{https://github.com/Arya-Honarpisheh/freq_iden_low_order}.}\footnote{This work has been submitted to the IEEE for possible publication. Copyright may be transferred without notice, after which this version may no longer be accessible.}
\end{abstract}


\vspace{-0.3cm}

\section{Introduction}
\label{sec:introduction}

\IEEEPARstart{S}{ystem} {identification is the process of constructing a mathematical model of a dynamical system from observed data, which may be collected in either the time or frequency domain \cite{keesman2011sysid}. Time domain methods work directly with input-output sequences and are advantageous for non-periodic excitations. Frequency domain methods \cite{mckelvey2002frequency, parrilo1998mixed_robust_iden, mckelvey1996subspace_freq} excel when system properties or control objectives are naturally expressed in terms of frequency-related quantities \cite{mashhadireza2022detection, mashhadireza2025novelty}, and are particularly appealing when a low-order model valid over a limited frequency band is sought \cite{mckelvey2000frequency_low_order}. From a computational standpoint, frequency domain methods are especially advantageous for systems with widely separated time constants: capturing fast transients requires a high sampling rate, while resolving slow modes demands a long observation horizon, together forcing large time-domain data records whose processing complexity scales cubic in the length of time-domain trajectories \cite{vanoverschee1996subspace}. Frequency domain methods sidestep this burden by operating only on the frequency points of interest, yielding substantial reductions in computational complexity.}


{From a control standpoint, low-order models---systems with small McMillan degree---are especially desirable. Modern robust control design methods yield controllers whose order matches that of the plant, so a high-order identified model directly inflates the complexity of the resulting controller. Beyond controller order, minimizing model order helps reduce overfitting, which in turn improves generalization. The standard approach to low-order system identification from time domain data} involves minimizing the rank of the Hankel matrix constructed from the system's impulse response \cite{fazel2013hankel__rank_sysid}. Other regularization based  methods include   \cite{pillonetto2016regularized_sysid, yilmaz2017randomized_pars_iden, khosravi2020lowcomplexity_sparse_estimation}. Use of frequency domain data was addressed in \cite{smith2014frequency_nuc_norm}, by transforming it to the time domain, thereby facilitating the formation of a Hankel matrix. {Finally, an  algorithm capable of handling both  time and frequency data, was proposed in \cite{sznaier2014fast_nnm}}.
 
In recent years there has been growing interest in non-asymptotic identification guarantees---bounds on the number of experiments required to achieve a given accuracy with high probability~\cite{tsiamis2023statistical_learning_control, simchowitz2018sysid_lti_mixing, chiuso2019sysid_learning,tu2024learning_many_trajectories}. Numerous
sample-complexity results now exist for LTI systems from time-domain
measurements~\cite{jedra2019sample_complexity,sarkar2019sample_complexity,zheng2020sample_complexity,oymak2021sample_complexity_ho_kalman,ziemann2022sample_complexity,Chatzikiriakos2024sample_complexity,fattahi2021sample_complexity},
and for partially observed systems~\cite{lee2022improved_rates_partial_obs,sattar2025finite_sample_part_bil}. For low-order identification specifically, sample complexity bounds have been
established in the time domain~\cite{fazel2022finite_sample_system_identification_nuclear_norm,cai2015robust_recovery_exp_signals_gaussian_hankel}. The work of~\cite{fazel2022finite_sample_system_identification_nuclear_norm} is closest in spirit to ours, recovering exponential signals via nuclear norm minimization of Hankel matrices: both share the nuclear norm regularization and minimum-gain argument, but differ in the data domain (time vs. frequency) and in the operator structure (Hankel vs. Loewner).
 
In the frequency domain, sample-complexity results remain sparse. The recent work of~\cite{tsiamis2024finitesamplefrequencydomain} establishes bounds for the classical Empirical Transfer Function Estimate (ETFE), but no sample-complexity analysis exists for regularization-based methods that explicitly promote low-order structure. The Loewner framework provides a natural vehicle for this purpose. Loewner matrices are central to interpolation-based system identification and model order reduction~\cite{karachalios2021loewner_framework_sysid_mor}. Their relevance comes from a fundamental rank property: a degree-$\kappa$ rational function induces a matrix of pairwise divided differences with rank at most $\kappa$, regardless of the number of frequency samples. Hence, infinitely many frequency-domain measurements can, in principle, be represented through a finite-rank object whose rank reflects the system order. Convex optimization methods exploiting this structure have been proposed~\cite{singh2020loewner_convex_low_order,honarpisheh2024loewner}, but sample-complexity bounds for such methods have not been available.

\emph{Contributions: }
This paper fills the gap by deriving the first
non-asymptotic sample-complexity bound for a regularization-based
frequency response identification algorithm. Specifically:
 
\begin{itemize}
  \item We propose a Loewner Nuclear Norm Minimization (LNNM) estimator that minimizes a quadratic data-fitting term regularized by the nuclear norm of the Loewner matrix---the frequency-domain analogue of Hankel nuclear norm regularization~\cite{fazel2022finite_sample_system_identification_nuclear_norm}---subject to a stability constraint enforced via a linear matrix inequality (LMI).
 
  \item We prove that, with probability at least $1-\delta$, the
    sampled-frequency identification error satisfies
    $\|{w}_\textrm{est} - {w}_\textrm{true}\|_\infty \leq \mathcal{O}(N^{-1/2})$, and derive
    an explicit finite-sample bound characterizing the dependence on the
    number of repeated experiments $N$, the number of frequency points
    $M$, the system order $\kappa$, and the noise level $\bar{\eta}$.  For
    a uniform frequency grid, the sample complexity grows as
    $\mathcal{O}\!\left(\sqrt{M \ln M / N}\right)$.  To the best of our
    knowledge, this is the first such bound for regularization-based
    frequency-domain algorithms.
 
  \item We extend the guarantee to all frequencies via rational
    interpolation and show that, under optimal tuning of $M$ and the
    frequency margin $\delta_m$, the overall $\mathcal{H}_\infty$
    identification error decays as $\mathcal{O}(N^{-1/3})$---matching the rate obtained for the ETFE in~\cite{tsiamis2024finitesamplefrequencydomain}.
 
  \item We validate the method and the predicted scaling laws through
    numerical experiments on SISO and MIMO systems, including Monte Carlo
    simulations that confirm the dependence on $N$, $M$, and $\bar{\eta}$.
    \item{We empirically demonstrate that integrating our frequency estimates with standard frequency-domain identification methods \cite{mckelvey1996subspace_freq,pintelon2012system} leads to system models that more accurately capture the unknown plant's behavior.}
\end{itemize}

The remainder of this paper is organized as follows. Section~\ref{sec:preliminaries} introduces the notation and Loewner background. Section~\ref{sec:main} states the problem formulation and our proposed solution. Section~\ref{sec:sample_complexity} develops the sample-complexity analysis. Section~\ref{sec:numerical_example} presents numerical experiments. Section~\ref{sec:conclusion} concludes the paper.

\section{Preliminaries}
\label{sec:preliminaries}

\subsection{Notation}
\label{sec:notation}

\begin{table}[ht]
\centering
    \begin{tabular}{@{} l l @{}}
    $\mathbb{Z}(\mathbb{Z}^+)$ & Set of integers (nonnegative integers) \\
    $\mathbb{R}^n (\mathbb{C}^n)$ & Set of real (complex) $n$-tuples \\
    $\mathbb{D}$ & Unit disk in the complex plane \\
    $\mathbb{D}_{\rho}$ & Disk with radius $\rho$ in the complex plane \\
    $\mathbb{R}^{n \times m} (\mathbb{C}^{n \times m})$ & Set of $n \times m$ real (complex) matrices \\
    $|z| (\arg(z))$ & Absolute value (argument) of the complex number $z$ \\
      $\bar{z}$ & Conjugate of the complex number $z$ \\
    $\Re(z) (\Im(z))$ & Real(Imaginary) part of the complex number $z$ \\
    $\bm{X}^*$ & Hermitian conjugate of the matrix $\bm{X}$ \\
    $\bm{X}^\dagger$ &pseudo-inverse of $\bm{X}$\\
    $\langle \bm{X}, \bm{Y} \rangle$ & inner product in the space of matrices: $\operatorname{Tr}(\bm{X}^* \bm{Y})$\\
    $\zero_n (\one_n)$ & The $n$-dimensional vector of all zeros(ones) \\
    $\bm{e}_k$ & The k-$th$ vector of the  standard basis of $\Rx{n}$ \\
    $\bm{I}_n$ & The $n \times n$ identity matrix \\
    $\bm{0}_{n \times m}$ & The $n \times m$ matrix of all zeros \\
    $\|\bm{x}\|_{p}$ & $\ell_{p}$-norm of vector $\bm{x}$ \\
    $\|\bm{X}\|_p$ & $p$-operator norm of matrix $\bm{X}$ \\
    $\|\bm{X}\|_*$ & Nuclear norm of matrix $\bm{X}$: sum of its singular values \\
    $\|\bm{X}\|_F$ & Frobenius norm of matrix $\bm{X}$ \\
    $\Tr(\bm{X})$ & Trace of matrix $\bm{X}$ \\
    $\mathcal{H}_\infty(\rho)$ &  Space of transfer functions with  bounded analytic con--\\& tinuation   inside the disk of radius $\rho > 1$, equipped with \\ & the norm \ $\|G \|_{\infty,\rho} = \sup_{|z|=\rho} |G(z)|$ \\
    $\mathcal{H}_\infty(\rho, K)$ & $\{G(z) \in \mathcal{H}_\infty(\rho) \colon \|G \|_{\infty,\rho} \leq K \}$ \\
 \end{tabular}
\vspace{-\baselineskip} 
\label{tab:notation}
\end{table}

For a vector $\bm{x}$, its $r^{\text{th}}$ element is denoted by $x_r$. For a matrix $\bm{X}$, the element in the $r$th row and $s$th column is represented by $x_{rs}$, while the $s$th column is denoted by $\bm{x}_s$.

Let $L: \mathcal{H} \to \mathcal{K}$ be a linear operator between Hilbert spaces $\mathcal{H}$ and $\mathcal{K}$. The adjoint operator of $L$ is denoted by $L^*$.

For a real-valued function $f: \mathcal{H} \to \mathbb{R}$ defined on the Hilbert space $\mathcal{H}$, the subdifferential of $f$ is denoted by $\partial f$.

Let $\mathcal{S}$ denote the set of discrete-time, causal, finite-dimensional LTI systems. Each system is characterized by its impulse response, denoted by the function $g:\mathbb{Z}^+ \rightarrow \mathbb{R}$. The corresponding transfer function is represented by $G:\mathbb{C} \rightarrow \mathbb{C}$ and is defined as $G(z) = \sum_{k=0}^\infty g(k) z^{k}$ \footnote{The transfer function defined herein differs from the standard notation in the literature, where the conventional transfer function is expressed as $G\left(\frac{1}{z}\right)$. For stable systems, the function $G(z)$ is analytic inside the unit disk. We retain this convention throughout as it allows the analyticity domain to coincide with the standard unit disk $\mathbb{D}$, simplifying the statement of Assumptions and stability constraints.}. A realization of a system in $\mathcal{S}$ is a state-space model:

\begin{equation*}
\begin{aligned}
    \bm{x}(k+1) &= \bm{A} \bm{x}(k) + \bm{B} \bm{u}(k) \\ 
    \bm{y}(k) &= \bm{C} \bm{x}(k) + \bm{D} \bm{u}(k),
\end{aligned}
\end{equation*}
with the state variable $\bm{x} \in \mathbb{R}^n$, such that
\begin{equation}
\label{eq:ss2tf}
    G(z) = \bm{C} \left( z^{-1} \bm{I}_n - \bm{A} \right)^{-1} \bm{B} + \bm{D}.
\end{equation}

\subsection{Loewner Matrices and Rational Interpolation}\label{sec:Loewner}
For ease of reference, below we recall some properties of Loewner matrices and their relation to rational interpolation.  A complete treatment can be found for instance in \cite{antoulas2017tutorial_loewner}. Given two sets of distinct points in the complex plane, $\{z^a_i\}_{i=1}^m$ and
$\{z^b_j\}_{j=1}^n$, along with corresponding function values $\{w^a_i\}$ and
$\{w^b_j\}$, the Loewner matrix $ \mathbf{L} \in \mathbb{C}^{n \times m}$
is defined \footnote{Throughout this paper, we use the symmetric partition $\bm{z}^{(a)} = \bm{z}$, $\bm{z}^{(b)} = \bar{\bm{z}}$, $\bm{w}^{(a)} = \bm{w}$, $\bm{w}^{(b)} = \bar{\bm{w}}$, yielding a square $M\times M$ Loewner matrix. This choice is standard for real-coefficient systems \cite{mayo2007framework_solution_generalized_realization}.}
\begin{equation}
\label{eq:loewner_matrix}
\mathbf{L} =    \begin{bmatrix}
        \frac{w^{(b)}_1 - w^{(a)}_1}{z^{(b)}_1 - z^{(a)}_1} & \cdots & \frac{w^{(b)}_1 - w^{(a)}_m}{z^{(b)}_1 - z^{(a)}_m} \\
        \vdots & \ddots & \vdots \\
        \frac{w^{(b)}_q - w^{(a)}_1}{z^{(b)}_n - z^{(a)}_1} & \cdots & \frac{w^{(b)}_n - w^{(a)}_p}{z^{(b)}_n - z^{(a)}_m}
    \end{bmatrix}.
\end{equation}

The Loewner matrix was introduced in the context of rational interpolation by Antoulas and
Anderson~\cite{Antoulas:1986}. Its rank carries direct system-theoretic meaning: it reveals the
\emph{order} of the underlying dynamical system. If $G(z)$ is a transfer function
of order $r$, then $\mathrm{rank}(\mathbf{L}) = r$, regardless of how many data
points were used to build it~\cite{Mayo:2007}.  Further,  an explicit expression of $G(z)$ in a Barycentric--Lagrange form can be explicitly constructed from the null space of $\mathbf{L}$:
\begin{equation}\label{eq:interpolant}
    G(z) = \frac{\sum_{i=1}^{m} \frac{w^a_i \, a_i}{z - z^a_i}}
                {\sum_{i=1}^{m} \frac{a_i}{z - z^a_i}}
\end{equation}
where $\mathbf{a} = [a_1, \ldots, a_{m}]^T$ is a vector in the right null space of $\mathbf{L}$. The Loewner matrix also admits a state-space factorization. It  can be shown that $\mathbf{L} = -\mathcal{O}\mathcal{R}$, where $\mathcal{O}$ and $\mathcal{R}$ are generalized observability and reachability matrices evaluated at the sample points~\cite{Mayo:2007}. This mirrors the classical factorization of the Hankel matrix $\mathbf{H} = \mathcal{O}\mathcal{R}$ in the time domain~\cite{Antoulas:2010}. The key advantage of $\mathbf{L}$ over $\mathbf{H}$ is that it captures the full infinite-horizon system behavior in a finite matrix \cite{Ionita:2013}. 

\subsection{Loewner Operator and its Adjoint} \label{sec:loewner_operator}
A closely related concept that will play a key role in establishing sampling complexity bounds is the \emph{complex Loewner operator} $\mathcal{L}_c(\bm{w}): \mathbb{C}^M \rightarrow \mathbb{C}^{M \times M}$: Given fixed frequencies $\bm{z}$, $\mathcal{L}_c(\bm{w}): \mathbb{C}^M \rightarrow \mathbb{C}^{M \times M}$ is the operator that, for each $\bm{{w}}$, returns the Loewner matrix $\mathbf{L}$, as defined in \eqref{eq:loewner_matrix}, where \(\bm{z}^{(a)},\bm{w}^{(a)}\) are taken to be $\bm{z},\bm{w}$ and \(\bm{z}^{(b)},\bm{w}^{(b)}\) are their conjugates $\bm{\bar{z}},\bm{\bar{w}}$. It is important to note that this operator is not linear over the field of complex numbers, as complex conjugation is inherently non-linear. However, when interpreted as a function defined over \(\mathbb{R}^{2M}\), the operator exhibits linearity. 
To formalize this perspective, we define the \emph{real Loewner operator} \(\mathcal{L}_r: \mathbb{R}^{2M} \to \mathbb{C}^{M \times M}\) as follows:
\begin{equation}
\label{eq:loewner_real}
  \mathcal{ L}_r\left(\begin{bmatrix} \bm{a} \\ \bm{b} \end{bmatrix}\right) = \mathcal{L}_c(\bm{a} + \bm{j} \bm{b}).
\end{equation}
Let $\bm{w} \in \ker(\mathcal{L}_r)$, then all the entries of $\mathcal{L}_r(\bm{w})$ are zero. From the requirement that all diagonal entries vanish, it follows that $w_k$ is real for each $k$. The condition that all off-diagonal entries are also zero further implies that $w_1 = \cdots = w_M$. Consequently, the real Loewner operator is not invertible, and its nontrivial kernel is spanned by $\ker(\mathcal{L}_r) = \begin{bmatrix} \bm{1}_M \\ \bm{0}_M \end{bmatrix}$. This non-invertibility  arises because adding a real constant to all entries of $\bm{w}$ leaves the Loewner matrix \eqref{eq:loewner_matrix} unchanged, introducing a degree of freedom in constructing a realization from $\mathcal{L}_c(\bm{w})$. For instance, Theorem~4.1 in \cite{mayo2007framework_solution_generalized_realization} constructs a realization in which the feedthrough matrix $\bm{D}$ is a free variable.

Denote the space of all Loewner matrices as $\mathbb{L}(M) = \mathcal{L}_c(\Cx{M}) = \mathcal{L}_r(\Rx{2M})$, and define the invertible version of $\mathcal{L}_r$ as the \emph{Loewner operator} $\mathcal{L}: \Rx{2M} \to \mathbb{L}(M)$. For any $\bm{X} \in \mathbb{L}(M)$, the inverse of $\bm{X}$ under $\mathcal{L}$ is defined as the unique element in $\mathcal{L}_r^{-1}(\bm{X})$ that has the minimum Euclidean norm. Thus, $\mathcal{L}^{-1}(\bm{X})$ is given by $\begin{bmatrix} \Re\bm{w} \\ \Im\bm{w} \end{bmatrix}$ such that $\sum \Re \bm{w} = 0$ since if $\begin{bmatrix} \Re\bm{w} \\ \Im\bm{w} \end{bmatrix} \in \mathcal{L}_r^{-1}(X)$, then for any $\lambda \in \mathbb{R}$, $\begin{bmatrix} \Re\bm{w} \\ \Im\bm{w} \end{bmatrix} + \lambda \begin{bmatrix} \bm{1}_M \\ \bm{0}_M \end{bmatrix}$ also belongs to $\mathcal{L}_r^{-1}(\bm{X})$.
Also of interests are  $\mathcal{L}^*$, the adjoint operator of $\mathcal{L}$ and its inverse $\mathcal{L}^{-*}$. Explicit expression for these operators are given in the Appendix~\ref{sec:adj_inv_loewner}.  

In what follows, with a slight abuse of notation we will denote $\mathcal{L}(\begin{bmatrix}\bm{a}^\top & \bm{b}^\top \end{bmatrix}^\top)$ with $\mathcal{L}(\bm{a} + \bm{j}\, \bm{b})$, and the same convention is applied to $\mathcal{L}^{-*}$. Additionally, if $ \mathcal{L}^{-1}(\bm{X}) = \begin{bmatrix} \bm{a}^\top & \bm{b}^\top \end{bmatrix}^\top,$ then it is identified with the complex number $\bm{a} + \bm{j}\bm{b}$. An analogous convention is adopted for $\mathcal{L}^*$.


\section{Low-Order Frequency-Domain Identification via Loewner Regularization}
\label{sec:main}

In this section we formally state the identification problem and propose a solution based on Loewner nuclear norm regularized least squares subject to a Nevanlinna--Pick stability constraint.

\subsection{Problem Statement}
\label{sec:problem_statement}

\begin{figure}[t]
    \centering
    \includegraphics[width=0.95\columnwidth]{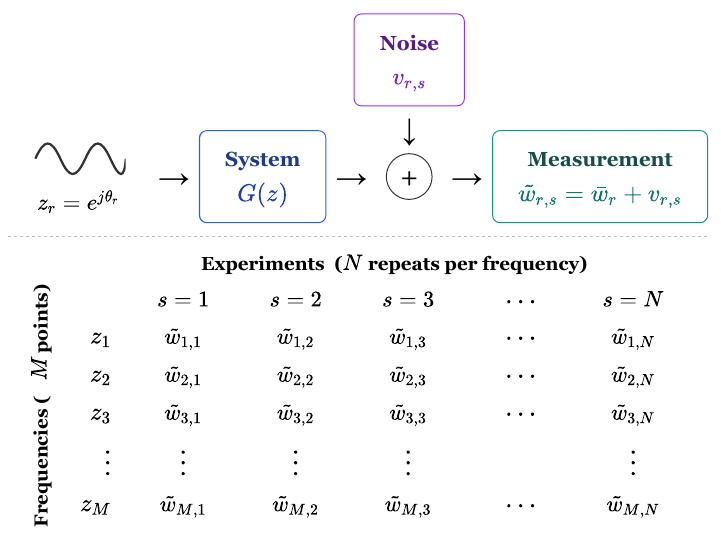}
    \caption{\emph{Data collection.} At each frequency $z_r = e^{j\theta_r}$,
    the system is excited sinusoidally and the steady-state amplitude/phase ratio
    is recorded. Repeating this over $N$ independent experiments at $M$ frequencies
    yields $T = MN$ noisy frequency-response measurements.}
    \label{fig:data_collection}
\end{figure}

We consider frequency-domain identification from repeated noisy measurements.
The dataset $\mathcal{D} = (\mathbf{z}, \widetilde{\mathbf{W}})$ consists of $N$
independent experiments conducted at $M$ frequency points
$\mathbf{z} = [z_1, \ldots, z_M]^\top$ on the upper unit semicircle:
\begin{equation}
\label{eq:frequency_data}
\widetilde{\mathbf{W}} = \bar{\mathbf{W}} + \mathbf{V}, \qquad
\widetilde{\mathbf{W}},\, \bar{\mathbf{W}},\, \mathbf{V} \in \mathbb{C}^{M \times N}
\end{equation}
where $\bar{\mathbf{W}}$ contains the true frequency responses and $\mathbf{V}$ is measurement noise. In each experiment $s = 1,\ldots,N$, the system is excited sinusoidally at all $M$ frequencies and the steady-state amplitude-phase ratio is recorded. The result is the $s$-th column $\tilde{\mathbf{w}}_s \in \mathbb{C}^M$ of $\widetilde{\mathbf{W}}$. This data collection procedure is illustrated in Figure~\ref{fig:data_collection}.

\emph{Goal: }
Given $\widetilde{\mathbf{W}}$, provide an estimation $\hat{\mathbf{w}} \in \mathbb{C}^M$ of the true sampled frequency response $\bar{\mathbf{w}} \in \mathbb{C}^M$ and, subsequently, a transfer function
$\hat{G}$ via realization. We are specifically interested in the \emph{nonasymptotic} regime: we seek the smallest number of experiments $N_{\mathrm{sc}}(\epsilon, \delta)$ such that, for all $N \geq N_{\mathrm{sc}}$,
\begin{equation}
\label{eq:sample_complexity}
    \mathrm{Prob}\!\left(\|\bm{\hat{w}} - \bm{\bar{w}}\|_\infty \leq \epsilon \right)
    \geq 1 - \delta.
\end{equation}
The quantity $N_{\mathrm{sc}}(\epsilon, \delta)$ is the \emph{sample complexity}
of the identification algorithm

\subsection{Identification Algorithm}\label{sec:algorithm}

We propose to recast the problem above into a constrained regularized \emph{Loewner Nuclear Norm Minimization (LNNM)}.

\begin{assumption}
\label{ass:apriori}
The system to be identified, $\bar{G}$,  belongs to $\mathcal{H}_\infty(\rho, K)$, where $\rho > 1$ and $K > 0$ are known.\end{assumption}
This is a standard assumption in interpolation-based identification \cite{parrilo1998mixed_robust_iden}.  If the information is not a priori available, $K$ can be directly estimated from the frequency domain data, while $\rho$ can be estimated from an impulse response experiment using the fact that the Markov parameters decay as $\rho^{-k}$ \cite{parrilo1998mixed_robust_iden}.

{Under this assumption, define the stable-trace feasible set
\begin{multline}
\label{eq:feasible_set}
\mathcal{W}_M(\rho,K) = \bigl\{\, \bm{w} \in \mathbb{C}^M : \exists\, G \in \mathcal{H}_\infty(\rho,K) \\
\text{s.t. } G(z_r) = w_r,\; r = 1,\dots,M \,\bigr\}.
\end{multline}
which collects all sampled frequency-response vectors compatible with at least one stable bounded transfer function.} The proposed LNNM estimator is obtained by solving
\begin{equation}
\label{eq:LNNM}
\begin{aligned}
\bm{\hat{w}} \in \operatorname*{arg\,min}_{\bm{w} \in \mathbb{C}^M} \quad &
\frac{\tau}{M}\norm{\mathcal{L}(\bm{w})}_* + \frac{1}{2} \sum_{s=1}^N \norm{\bm{w} - \bm{\tilde{w}}_s}_2^2 \\
\text{subject to} \quad &
\begin{bmatrix}
Q & \frac{1}{K} W \\[4pt]
\frac{1}{K} W^* & Q^{-1}
\end{bmatrix} \succeq 0, \\
& W = \operatorname{diag}(\bm{w}), \\
& Q_{ij} = \dfrac{1}{1 - \frac{z_i}{\rho}\frac{\bar{z}_j}{\rho}}, \qquad i,j = 1,\ldots,M .
\end{aligned}
\end{equation}
Here $\tau > 0$ is a regularization parameter, $ \bm{\hat{w}}$ denotes the frequency response of the identified system at the frequency points $\bm{z}$, and $\mathcal{L}(\bm{{w}})$ is the Loewner matrix obtained by using a symmetric partitioning of the data, namely $\bm{z}^{(a)} = \bm{z}$, $\bm{z}^{(b)} = \bar{\bm{z}}$, $\bm{w}^{(a)} = \bm{w}$, and $\bm{w}^{(b)} = \bar{\bm{w}}$, yielding a square $M\times M$ Loewner matrix. The constraint in \eqref{eq:LNNM}, arising from Nevanlinna Pick interpolation, guarantees that $\bm{\hat{w}}$ is consistent with the a-priori information, in other words $\bm{\hat{w}} \in \mathcal{W}_M(\rho,K)$ \cite{parrilo1998mixed_robust_iden}

\begin{remark}
\label{rem:nuclear_norm_motivation}
The objective in  \eqref{eq:LNNM} balances fidelity to the data against the order of the identified system. The term
$\norm{{\mathcal{L}(\bm{w})}}_*$ acts as a surrogate for rank  \cite{fazel2001rank_minimization_min_order}, thus favoring low-rank solutions while preserving computational tractability. In turn, this promotes low order systems, since as noted in section \ref{sec:Loewner}, the order of the (rational)  system that generated the data is precisely
 the rank of the Loewner matrix \cite{mayo2007framework_solution_generalized_realization}.
 \end{remark}

\begin{remark}
\label{rem:LNNM_remark}
Compared to other convex optimization-based methods for identifying low-order systems in the frequency domain \cite{singh2020loewner_convex_low_order, honarpisheh2024loewner}, the proposed approach does not impose agreement with the measured data as hard interpolation constraints. Instead, the noisy measurements enter only through the quadratic loss term, while stability is enforced separately via the LMI constraint in \eqref{eq:LNNM}. The nuclear norm of the Loewner matrix scaled by the number of frequency points, $\frac{\| \mathcal{L}(\bm{w}) \|_*}{M}$ converges to the trace norm of the system i.e. the sum of the singular values of the Hankel operator (Theorem~1 in \cite{honarpisheh2024loewner}). 
\end{remark}

\begin{remark} The optimal solution to \eqref{eq:LNNM} lies on the boundary of the feasible set, hence the Pick matrix $P = Q-\frac{1}{K^2}WQW^* \succeq 0$. Thus, the transfer function that interpolates the points $(\bm{z},\bm{\hat{w}})$ is unique (\cite{garnett1981bounded}, Corollary 2.3). A state space realization of this interpolant given in terms of $P$ can be found for instance using the formulas in \cite{parrilo1998mixed_robust_iden}.
\end{remark}

\section{Sample Complexity Analysis}
\label{sec:sample_complexity}

In this section we present the main result of the paper: a sample complexity analysis showing that, with high probability, the solution of \eqref{eq:LNNM} satisfies $ \| \bm{\hat{w}} - \bm{\bar{w}} \|_\infty \leq \mathcal{O}(N^{-1/2})$, which aligns with the general expectations for identification errors.  The analysis in this section is developed for single-input-single-output (SISO) systems. Extending this analysis to the MIMO case  requires non-trivial extensions of the concentration and Loewner-operator bounds to the matrix-valued setting and is left for future work. Nevertheless, a  numerical demonstration of the MIMO extension is provided in Section~\ref{sec:mimo_identification}.

We make the following assumptions throughout the rest of the paper:


\begin{assumption}
\label{ass:noise}
The noise has zero mean, i.e., $\mathbb{E}(v_{rs}) = 0$, is independent across experiments and frequencies (it does not need to be identically distributed), and is almost surely bounded such that $\mathbb{P}(|v_{rs}| \leq \eta_r) = 1$, where the bound $\eta_r$ depends on the frequency. Consequently, $v_{rs}$ is a sub-Gaussian random variable with variance proxy $\eta_r^2$, which means its variance is bounded by $\eta_r^2$. We denote by $\bar{\eta} = \max_{r \leq M} (\eta_r)$ the worst-case noise bound across all frequencies.
\end{assumption}

\begin{figure}[h]
    \centering
    \includegraphics[width=0.55\linewidth]{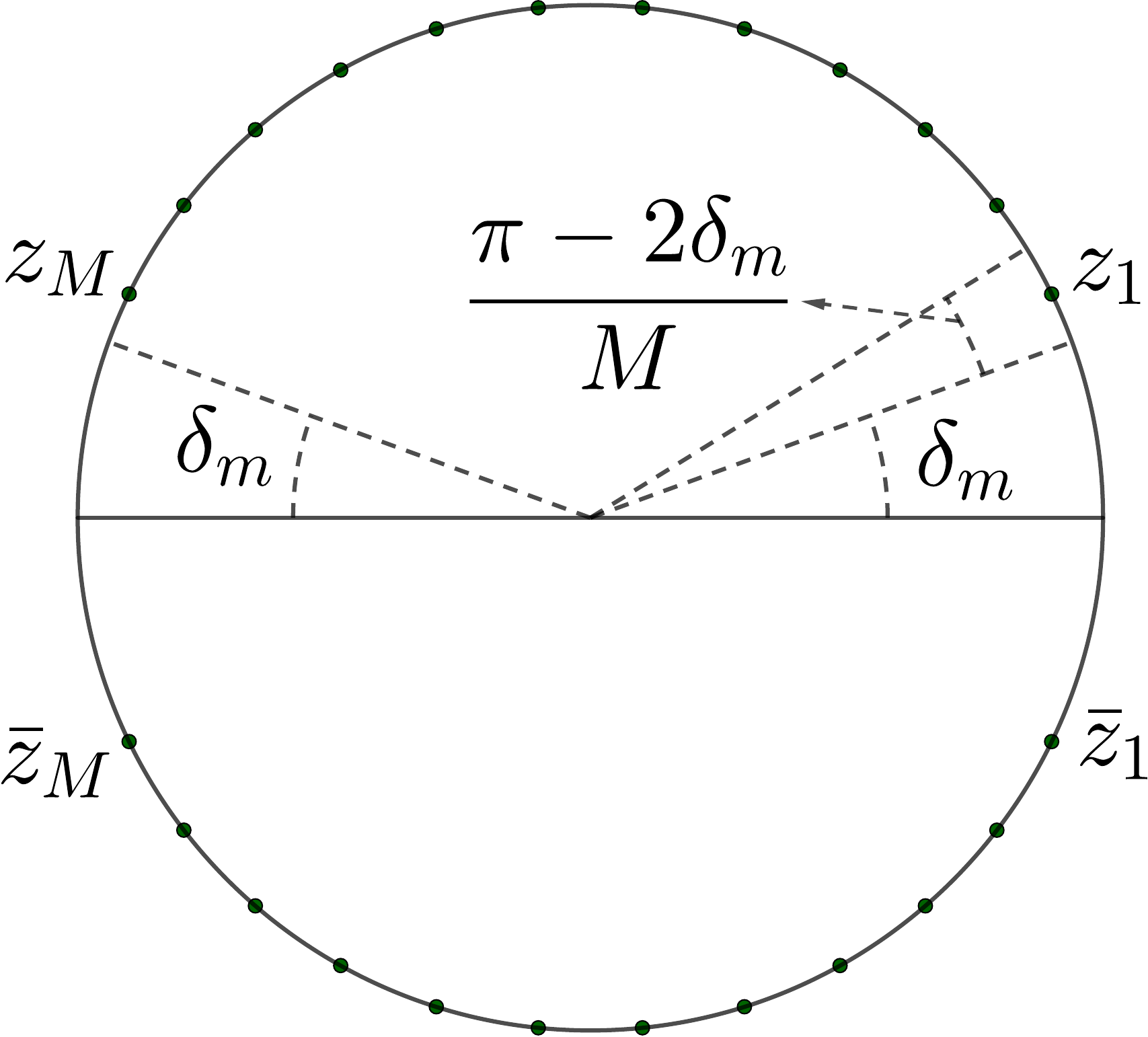}
    \caption{\emph{Frequency Points $\bm{z}$:} Illustration of the equally spaced special case on the upper half of the unit circle with margin $\delta_m$ from $0$ and $\pi$.}
    \label{fig:circle}
\end{figure}

\begin{assumption}
\label{ass:z_margin}
The frequency points $z_r = e^{\bm{j}\theta_r}$ lie on the upper half of the unit circle, where $\theta_1 < \cdots < \theta_M$ are the midpoints of an arbitrary partition $\delta_m = p_0 < p_1 < \cdots < p_M = \pi - \delta_m$ of $[\delta_m,\pi-\delta_m]$, with subinterval widths $h_r = p_r - p_{r-1}$. Denote $h_{\max} = \max_r h_r$ and $h_{\min} = \min_r h_r$. The parameter $0 < \delta_m < \pi/2$ is the frequency margin. The equally spaced case $h_r = (\pi-2\delta_m)/M$ for all $r$ is a special case shown in Figure~\ref{fig:circle}.
\end{assumption}

\begin{remark}
    Although our proposed method can be readily applied to the identification of LTI systems without these technical assumptions, they are essential for the derivation of the sample complexity bounds.
    \begin{itemize}
        \item Assumption~\ref{ass:noise} is a standard bounded noise assumption. We believe that this assumption is not overly restrictive, as the analysis can be modified to accommodate alternative noise distributions (e.g., Gaussian or colored noise) by leveraging concentration inequalities different from Theorem~\ref{thm:matrix_hoeffding}.
        \item Assumption~\ref{ass:z_margin} is a new assumption introduced in this paper; it requires that there exists a margin from $0$ and $\pi$ for the frequency points $\bm{z}$. Although $\delta_m$ appears in the sample complexity bound, for any finite number of $\bm{z}$, as long as $1$ and $-1$ are not included in $\bm{z}$, such a positive $\delta_m$ can be found. Conversely, if $1$ or $-1$ are present in $\bm{z}$, they may be disregarded. Technically, the frequency response at $1$ corresponds to the DC gain of the system. In practice, it is not used to construct the Loewner matrix but is instead employed later to determine the input-to-output matrix $D$ as previously discussed.
    \end{itemize}
\end{remark}

We are now ready to state the main theorem, which provides a finite-sample bound on the sampled-frequency estimation error. Its proof is postponed until after the required preliminary results are established.

\begin{theorem}
\label{thm:sample_complexity_finite_sample}
    Let the experimental frequency data be given as in \eqref{eq:frequency_data} and denote by $\bm{\hat{w}}$ the sampled frequency-response vector obtained by solving the optimization problem \eqref{eq:LNNM}. Provided  that Assumptions \ref{ass:apriori}, \ref{ass:noise}, and \ref{ass:z_margin} hold, and that \(h_{\max}\) and \(\delta_m\) are small enough (as specified in the proof of Lemma~\ref{lem:L_F_to_u_2_norm_bound}), then, for the optimal choice of $\tau$ as 
    \begin{equation}
    \label{eq:tau_optimal}
        \tau^* = (1+\sqrt{2})\,\frac{16\bar{\eta}\sqrt{-N\ln\sin\delta_m\,\ln(M/\delta)}}{h_{\min}}
    \end{equation}
    the sampled-frequency estimation error satisfies,  with probability at least $1 - \delta$,
    \begin{equation}
    \label{eq:finite_sample_bound}
    \begin{aligned}
    \| \bm{\hat{w}} - \bm{\bar{w}} \|_\infty &\leq 192(1+\sqrt{2})^2\sqrt{2\kappa} \\ & \sqrt{1+\frac{4K}{\rho-1}}\,\bar{\eta}\,
    \frac{(-\ln\sin\delta_m)\sqrt{h_{\max}}}{\sqrt{\pi}\,h_{\min}^2\,M}
    \sqrt{\frac{\ln(M/\delta)}{N}}
    \end{aligned}
    \end{equation}
    in which $\kappa$ is the true order of the system.
\end{theorem}

The low-order structure of the true system enters this  bound through $\sqrt{\kappa}$,
so systems with smaller McMillan degree $\kappa$ yield tighter guarantees. When $\kappa = 1$, $M$ can be chosen to meet the interpolation requirement without the bound deteriorating due to system complexity. In practice, one need not avoid large $M$ when $\kappa$ is small: provided $M \ln M / N$ remains small, a finer frequency grid improves frequency resolution at no additional complexity cost.

For fixed \(M\) and \(\delta_m\), the mesh-dependent factor $\frac{\sqrt{h_{\max}}}{M h_{\min}^2}$ is minimized, up to constants, by a uniform grid. In particular, for the uniform grid \(h_{\max}=h_{\min}=(\pi-2\delta_m)/M\), this factor scales as \(\mathcal{O}(\sqrt M)\). Hence the nonuniform theory allows flexibility in choosing the frequency points, but the most favorable \(M\)-dependence of the bound is obtained when the grid is not highly clustered.


\begin{remark}
\label{rem:comp_ave}
A natural unregularized baseline is the pointwise averaged estimator $\hat{w}^{\rm avg}_r = \frac{1}{N}\sum_{s=1}^N \tilde{w}_{r,s}$. Under Assumption~\ref{ass:noise}, Hoeffding's inequality applied to the real and imaginary parts separately, followed by a union bound over $M$ frequencies, gives, with probability at least $1-\delta$, $\|\hat{\bm{w}}^{\rm avg} - \bar{\bm{w}}\|_\infty \leq
2\bar{\eta}\sqrt{\ln(2M/\delta)/N}$. Compared to \eqref{eq:finite_sample_bound}, the LNNM bound carries an
additional $\sqrt{M}$ factor, which does not arise from the noise concentration step but from the minimum-gain bound on the Loewner operator in \eqref{eq:phi_bound} specific to our problem settings.
\end{remark}

\begin{remark}
\label{rem:total_samples}
Writing $T = MN$ for the total number of measurements, the bound $\mathcal{O}(\sqrt{M/N})$ becomes $\mathcal{O}(M/\sqrt{T})$. For a fixed total measurement budget $T$, increasing $M$ reduces $N = T/M$ and makes the bound worse, reflecting the tradeoff between frequency resolution and per-frequency averaging. This $\mathcal{O}(M/\sqrt{T})$ scaling differs from ETFE-type bounds, which scale as $\mathcal{O}(\sqrt{M/T})$ \cite{tsiamis2024finitesamplefrequencydomain}: in the ETFE setting, $T$ counts time-domain samples and $M$ is a post-processing choice, whereas here $M$ is part of the experimental design --- the system is physically excited at $M$ frequencies with $N$ repetitions each. The additional $\sqrt{M}$ factor in \eqref{eq:finite_sample_bound} relative to ETFE therefore reflects a genuine difference in problem structure, not a suboptimality of the analysis.
\end{remark}


Theorem~\ref{thm:sample_complexity_finite_sample} controls the identification error only at the sampled frequency points. We next state a direct consequence extending this guarantee to all frequencies, assuming that the extension to all frequencies is done using an interpolation-based method \cite{chen1995Hinf_interpol_sysid}.

\begin{corollary}
\label{cor:all_freq}
    If an interpolation based algorithm is used to identify  $\hat{G}$ from its sampled  frequency response obtained solving \eqref{eq:LNNM} and assumptions \ref{ass:apriori}, \ref{ass:noise}, and \ref{ass:z_margin} hold, then with probability at least $1 - \delta$ the following bound holds
    \begin{equation}
    \label{eq:final_bound}
        \| \hat{G} - \bar{G} \|_\infty \leq 2 \frac{\epsilon + K \psi(\bar{\delta})}{1 + \frac{\epsilon}{K} \psi(\bar{\delta})}
    \end{equation}
    where  $\epsilon$ is the RHS of \eqref{eq:finite_sample_bound} and 
    \begin{equation*}
        \psi(\bar{\delta}) = \frac{2 \rho \sin (\frac{\bar{\delta}}{2})}{\sqrt{(\rho^2 - 1)^2 + 4 \rho^2\sin^2(\frac{\bar{\delta}}{2})}}.
    \end{equation*}
    $\bar{\delta}$ is the maximum arc distance between each pair in $\begin{bmatrix} \bm{z}^\top \bm{\bar{z}}^\top \end{bmatrix}^\top$.
\end{corollary}
\begin{proof}
    The proof is a direct consequence of Theorem 5.2 in \cite{chen1995Hinf_interpol_sysid} and Theorem~\ref{thm:sample_complexity_finite_sample}.
\end{proof}

The number of frequency points $M$ and the frequency margin $\delta_m$ jointly influence the identification error in \eqref{eq:final_bound} through both $\epsilon$ and $\psi(\bar{\delta})$. The term $\psi(\bar{\delta})$ represents an interpolation error: how well the identified model at the $M$ grid points approximates $\bar{G}$ over all frequencies. Since $\psi(\bar{\delta}) = \mathcal{O}(\bar{\delta})$ for small $\bar{\delta}$, and $\bar{\delta} = \max\{2\delta_m + h_{1}, h_{\max} \}$, driving $\psi(\bar{\delta})$ to zero requires not only $h_{\max} \to 0$ but also $\delta_m\to 0$. The natural choice is $\delta_m = \mathcal{O}(1/M)$, meaning the boundary margin scales with the grid spacing, corresponding to a uniform grid over the full frequency range. Under this choice, $\psi(\bar{\delta}) = \mathcal{O}(1/M)$ and, for a uniform grid where $h_{\max} = h_{\min} = (\pi-2\delta_m)/M$, ignoring logarithmic factors, $\epsilon = \mathcal{O}(\sqrt{M/N})$. Balancing both $\epsilon$ and $\psi(\bar{\delta})$ yields the optimal tuning $M = \mathcal{O}(N^{1/3})$ and $\delta_m = \mathcal{O}(N^{-1/3})$, under which both terms decay as $\mathcal{O}(N^{-1/3})$, giving an overall $\mathcal{H}_\infty$ identification error of $\mathcal{O}(N^{-1/3})$. This matches the rate obtained for the ETFE in Theorem~3 of \cite{tsiamis2024finitesamplefrequencydomain}.

It remains to prove Theorem~\ref{thm:sample_complexity_finite_sample}.
We decompose the proof into several intermediate steps as follows.

\begin{itemize}
    \item Section~\ref{sec:initial_bound}: Derive a preliminary sampled-frequency error bound from the optimality conditions of \eqref{eq:LNNM}, reducing the problem to controlling a minimum-gain term and a transformed noise term.
    \item Section~\ref{sec:minimum_gain}: Lower bound the minimum-gain term by exploiting the low-order structure of the true system together with the stability constraint.
    \item Section~\ref{sec:noise_term}: Bound the transformed noise term with high probability using a concentration argument.
    \item Section~\ref{sec:sampled_frequency_bound}: Combine these ingredients and optimize \(\tau\) to prove Theorem~\ref{thm:sample_complexity_finite_sample}.
\end{itemize}

\subsection{Initial Bound on the Identification Error}
\label{sec:initial_bound}

In this section, we first derive a preliminary bound on sample complexity (Lemma~\ref{lem:finit_sample_bound}), which depends on a parameter called the minimum gain, $\phi$, and includes a term that depends on the noise. There is no analytical solution to \eqref{eq:LNNM}, and thus it is not possible to explicitly compute $\bm{\hat{w}}$ to directly bound the sampled-frequency identification error $\| \bm{\hat{w}} - \bm{\bar{w}} \|_\infty$ by plugging in $\bm{\hat{w}}$. Instead, we derive the bound directly from the basic optimality inequality, which holds because $\bm{\bar{w}} \in \mathcal{W}_M(\rho,K)$ is a feasible point for \eqref{eq:LNNM}. If we run only one experiment at each frequency ($N=1$) and replace the Loewner operator $\LL$ in \eqref{eq:LNNM} with the diagonal embedding $\bm{w}\mapsto\operatorname{diag}(\bm{w})$, the nuclear norm reduces to $\|\operatorname{diag}(\bm{w})\|_* = \|\bm{w}\|_1$ and \eqref{eq:LNNM} becomes the standard $\ell_1$-penalized least squares problem. In this case, the objective is to recover a sparse vector from its noisy measurements (the denoising problem). Motivated by this observation, we extend the approach in \cite{bhaskar2013atomic_norm_denoising_line_spectral_estimation} to derive the sample complexity bounds for our setting. The next lemma establishes a preliminary upper bound on the sample complexity of our proposed method. This bound quantifies the identification error at a finite set of frequency points, specifically $\bm{z}$.

\begin{lemma}
\label{lem:finit_sample_bound}
Let the experimental frequency data be given as in \eqref{eq:frequency_data}, and let Assumption~\ref{ass:apriori} hold. Denote by $\bm{\hat{w}}$ the sampled frequency-response vector obtained by solving the optimization problem \eqref{eq:LNNM}. Then, the sampled-frequency identification error satisfies
\begin{equation}
\label{eq:finite_sample_general_bound}
\begin{aligned}
\norm{\bm{\hat{w}} - \bm{\bar{w}}}_\infty &\leq \norm{\bm{\hat{w}} - \bm{\bar{w}}}_2 \leq \\
&\frac{2}{\phi} \left( \frac{\tau}{MN} + \frac{\| \sum_{s=1}^N L^{-*}(\bm{v}_s) \|_2}{N} \right).
\end{aligned}
\end{equation}
where \(\phi\) is the \emph{stable minimum gain} of the Loewner operator over
\[
\mathcal T_{\bm{\bar w}}^{\rm st}:=T_{\bm{\bar w}}\cap \mathcal W_M(\rho,2K),
\]
with
\[
T_{\bm{\bar{w}}}
=
\left\{
\bm{u} \in \mathbb{C}^M :
\norm{L(\bm{\bar{w}}+\bm{u})}_*
\leq
\norm{L(\bm{\bar{w}})}_*
+
\alpha\norm{L(\bm{u})}_*
\right\},
\]
with
\begin{equation}
\label{eq:alpha}
\alpha =
\frac{M}{\tau}
\left\|
\sum_{s=1}^N L^{-*}(\bm{v}_s)
\right\|_2 .
\end{equation}
and
\begin{equation}
\label{eq:minimum_gain}
\phi :=
\inf_{\substack{\bm u\in \mathcal T_{\bm{\bar w}}^{\rm st}\\ \bm u\neq 0}}
\frac{\norm{\bm u}_2}{\norm{L(\bm u)}_*}.
\end{equation}

\end{lemma}

\begin{proof}
Set $\bm{u} = \bm{\hat{w}} - \bm{\bar{w}}$. Since Assumption~\ref{ass:apriori} gives $\bm{\bar{w}} \in \mathcal{W}_M(\rho,K)$ and $\bm{\hat{w}}$ is feasible for \eqref{eq:LNNM}, the basic feasibility inequality holds:
\begin{equation}
\label{eq:basic_ineq}
\begin{aligned}
&\frac{\tau}{M}\norm{L(\bm{\hat{w}})}_* + \frac{1}{2}\sum_{s=1}^N \norm{\bm{\hat{w}} - \bm{\tilde{w}}_s}_2^2 \\
&\qquad \leq \frac{\tau}{M}\norm{L(\bm{\bar{w}})}_* + \frac{1}{2}\sum_{s=1}^N \norm{\bm{\bar{w}} - \bm{\tilde{w}}_s}_2^2.
\end{aligned}
\end{equation}
Substituting $\bm{\tilde{w}}_s = \bm{\bar{w}} + \bm{v}_s$ and using $\norm{\bm{u} - \bm{v}_s}_2^2 - \norm{\bm{v}_s}_2^2 = \norm{\bm{u}}_2^2 - 2\Re\langle \bm{u}, \bm{v}_s \rangle$, we obtain
\begin{equation}
\label{eq:basic_ineq_expanded}
\frac{N}{2}\norm{\bm{u}}_2^2 \leq \Re\!\left\langle \bm{u}, \sum_{s=1}^N \bm{v}_s \right\rangle + \frac{\tau}{M}\bigl(\norm{L(\bm{\bar{w}})}_* - \norm{L(\bm{\hat{w}})}_*\bigr).
\end{equation}
By the definition of the adjoint operator and nuclear/spectral duality:
\begin{equation}
\label{eq:inner_adj_dual_ineq}
\begin{aligned}
&\Re\!\left\langle \bm{u}, \sum_s \bm{v}_s \right\rangle = \Re\!\left\langle L(\bm{u}),\, L^{-*}\!\left(\sum_s \bm{v}_s\right) \right\rangle \\
&\qquad \leq \norm{L(\bm{u})}_* \left\| \sum_s L^{-*}(\bm{v}_s) \right\|_2.
\end{aligned}
\end{equation}
Substituting this bound into \eqref{eq:basic_ineq_expanded} and applying the reverse triangle inequality $\norm{L(\bm{\bar{w}})}_* - \norm{L(\bm{\hat{w}})}_* \leq \norm{L(\bm{u})}_*$, we obtain
\begin{equation}
\label{eq:before_min_gain}
\frac{N}{2}\norm{\bm{u}}_2^2 \leq \left( \frac{\tau}{M} + \left\| \sum_{s=1}^N L^{-*}(\bm{v}_s) \right\|_2 \right) \norm{L(\bm{u})}_*.
\end{equation}
If $\bm u=0$, the claim is trivial. Otherwise, dividing both sides by
$\frac{N}{2}\norm{\bm u}_2$ yields
\begin{equation}
\label{eq:fs_error_bound_no_constant}
\norm{\bm{u}}_2 \leq \frac{2\norm{L(\bm{u})}_*}{N\norm{\bm{u}}_2} \left( \frac{\tau}{M} + \left\|\sum_{s=1}^N L^{-*}(\bm{v}_s)\right\|_2 \right).
\end{equation}
For the second part of the proof, we establish cone membership of $\bm{u}$. Dropping the nonnegative term $\frac{N}{2}\norm{\bm{u}}_2^2$ from \eqref{eq:basic_ineq_expanded} and applying \eqref{eq:inner_adj_dual_ineq} gives
\begin{equation*}
\tau \frac{\norm{L(\bm{\hat{w}})}_*}{M} \leq \tau \frac{\norm{L(\bm{\bar{w}})}_*}{M} + \| L(\bm{u}) \|_* \left\| \sum_{s=1}^N L^{-*} (\bm{v}_s) \right\|_2,
\end{equation*}
which leads to
\begin{equation*}
\norm{L(\bm{\hat{w}})}_* \leq \norm{L(\bm{\bar{w}})}_* + \frac{M \| \sum_{s=1}^N L^{-*} (\bm{v}_s) \|_2}{\tau} \| L(\bm{u}) \|_*.
\end{equation*}
This implies $\bm{u} \in T_{\bm{\bar{w}}}$ with $\alpha$ as in \eqref{eq:alpha}. 
Moreover, since $\bm{\hat w}$ is feasible for \eqref{eq:LNNM}, $\bm{\hat w}\in\mathcal W_M(\rho,K)$, while Assumption~\ref{ass:apriori} gives $\bm{\bar w}\in\mathcal W_M(\rho,K)$. Hence $G_u:=G_{\hat w}-\bar G\in\mathcal H_\infty(\rho,2K)$ interpolates $\bm u=\bm{\hat w}-\bm{\bar w}$, so $\bm u\in\mathcal W_M(\rho,2K)$. Therefore $\bm u\in\mathcal T_{\bm{\bar w}}^{\rm st}$. Since $\phi$ is defined in \eqref{eq:minimum_gain} as the infimum of $\|\bm u\|_2/\|L(\bm u)\|_*$ over all nonzero $\bm u\in\mathcal T_{\bm{\bar w}}^{\rm st}$, we have $\|L(\bm u)\|_*\leq \|\bm u\|_2/\phi$. 
Substituting this inequality into \eqref{eq:fs_error_bound_no_constant} completes the proof.

\end{proof}

Note that ensuring $\alpha < 1$ is essential for the validity of the stable minimum gain argument. When $\alpha = 0$, the cone $T_{\bar{\bm{w}}}$ contains exactly the directions along which the nuclear norm of the Loewner matrix decreases, which is desirable since our goal is to recover a low-order system. However, deriving a bound for the case $\alpha = 0$ is not possible here, as the optimization in \eqref{eq:LNNM} jointly minimizes not only the nuclear norm of the Loewner matrix but also the squared distance between the measured and estimated frequency responses. For this reason, the cone must be widened by allowing $\alpha > 0$. Nonetheless, if $\alpha = 1$, the inequality becomes trivial and the directions in $T_{\bar{\bm{w}}}$ no longer represent descent directions for minimizing the nuclear norm. Thus, keeping $\alpha$ strictly less than one is important. Later, in the proof of Theorem~\ref{thm:sample_complexity_finite_sample}, we will show that choosing $\tau$ optimally ensures $\alpha < 1$.

\begin{remark}
\label{rem:bias_variance}
An examination of \eqref{eq:finite_sample_general_bound} reveals a conceptual resemblance to the classical bias-variance tradeoff in statistics \cite{hastie2009esl}. In this formulation, the first term represents the bias, regulated by the parameter $\tau$, which steers the model toward low-order systems. The second term is analogous to the variance, arising from the presence of noise in the data.
\end{remark}

\subsection{Bounding the Stable Minimum Gain}
\label{sec:minimum_gain}

Next, we derive a lower bound on the stable minimum gain of the Loewner operator over \(\mathcal T_{\bm{\bar w}}^{\rm st}\). To this end, let Assumptions \ref{ass:apriori} and \ref{ass:z_margin} hold. Consider the singular value decomposition of the Loewner matrix constructed from the true frequency response:
\begin{equation*}
\label{eq:svd_L_wdot}
    L(\bm{\bar{w}}) =
    \begin{bmatrix}
        \bm{U}_1 & \cdots
    \end{bmatrix}
    \begin{bmatrix}
        \bm{\Sigma}_1 & \bm{0} \\
        \bm{0} & \bm{0}
    \end{bmatrix}
    \begin{bmatrix}
        \bm{V}_1 & \cdots
    \end{bmatrix}^* = 
    \bm{U}_1 \bm{\Sigma}_1 \bm{V}_1^*,
\end{equation*}
where $\bm{\Sigma}_1$ is of dimension $\kappa$, representing the true order of the system.

We adopt the projection technique widely used in the literature~\cite{recht2010guaranteed, candes2012exact_matrix_comp_convex, cai2015robust_recovery_exp_signals_gaussian_hankel, bhaskar2013atomic_norm_denoising_line_spectral_estimation, fazel2022finite_sample_system_identification_nuclear_norm}. Define the subspace
\[
T = \left\{ \bm{U}_1 \bm{X} + \bm{Y} \bm{V}_1^* \; : \; \bm{X} \in \mathbb{C}^{\kappa \times M},\, \bm{Y} \in \mathbb{C}^{M \times \kappa} \right\},
\]
which corresponds to the tangent space, at $L(\bm{\bar{w}})$, to the manifold of matrices of rank $\kappa$. The orthogonal subspace to $T$ is
\[ 
T^\perp = \left\{ \bm{Z} \in \mathbb{C}^{M \times M} \; : \; \bm{U}_1^* \bm{Z} = \bm{0}_{\kappa \times M}, \, \bm{Z} \bm{V}_1 = \bm{0}_{M \times \kappa} \right\}.
\]
The projection operators onto these subspaces are given by
\begin{align*}
\mathcal{P}(\bm{X}) &= \bm{U}_1 \bm{U}_1^* \bm{X} + \bm{X} \bm{V}_1 \bm{V}_1^* - \bm{U}_1 \bm{U}_1^* \bm{X} \bm{V}_1 \bm{V}_1^*, \\
\mathcal{P}^\perp(\bm{X}) &= (\bm{I}_M - \bm{U}_1 \bm{U}_1^*) \bm{X} (\bm{I}_M - \bm{V}_1 \bm{V}_1^*).
\end{align*}

Let $\bm{u} \in \mathcal T_{\bm{\bar w}}^{\rm st}$. Since $\bm u\in T_{\bm{\bar w}}$,
\begin{equation*}
    \|L(\bm{\bar{w}} + \bm{u})\|_* \leq \|L(\bm{\bar{w}})\|_* + \alpha \|L(\bm{u})\|_*.
\end{equation*}
By the linearity of the Loewner operator, this simplifies to
\begin{equation}
\label{eq:Lwdot_Lu_leq_Lwdot_alpha}
    \|L(\bm{\bar{w}}) + L(\bm{u})\|_* \leq \|L(\bm{\bar{w}})\|_* + \alpha \|L(\bm{u})\|_*.
\end{equation}
On the other hand, decompose $L(\bm{u})$ as $L(\bm{u}) = \mathcal{P}(L(\bm{u})) + \mathcal{P}^\perp(L(\bm{u}))$, and write the singular value decomposition $\mathcal{P}^\perp(L(\bm{u})) = \bm{U}_2 \bm{\Sigma}_2 \bm{V}_2^* \in T^\perp$. From the characterization of the subdifferential of the nuclear norm~\cite{watson1992characterization_subdifferentil_matrix_norm}, we have
\[
\partial \|\cdot\|_*(L(\bm{\bar{w}})) = \left\{ \bm{U}_1 \bm{V}_1^* + \bm{Z} \;:\; \bm{Z} \in T^\perp,\; \|\bm{Z}\|_2 \leq 1 \right\}.
\]
Let $\bm{H} = \bm{U}_2 \bm{V}_2^*$. Then $\bm{H} \in T^\perp$ and $\|\bm{H}\|_2 = 1$. Therefore, $\bm{U}_1 \bm{V}_1^* + \bm{H}$ is a valid subgradient at $L(\bm{\bar{w}})$. The form of this subdifferential has the following interpretation: $\bm{U}_1\bm{V}_1^*$ is the subgradient at any full-rank matrix and captures the contribution of the $\kappa$ nonzero singular values, while $\bm{Z} \in T^\perp$ with $\|\bm{Z}\|_2 \leq 1$ accounts for the free directions in the null space of $L(\bm{\bar{w}})$ where the nuclear norm is not differentiable. Recall that for any subgradient $\bm{G} \in \partial\|\cdot\|_*(L(\bm{\bar{w}}))$, the subgradient inequality gives $\|L(\bm{\bar{w}}) + \bm{\Delta}\|_* \geq \|L(\bm{\bar{w}})\|_* + \langle \bm{G}, \bm{\Delta}\rangle$ for any $\bm{\Delta}$. Taking $\bm{G} = \bm{U}_1\bm{V}_1^* + \bm{H}$ and $\bm{\Delta} = L(\bm{u})$ yields
\begin{equation}
\label{eq:Lw_Lu_geq_Lw_innprod_UVH_Lu}
\begin{aligned}
    &\|L(\bm{\bar{w}}) + L(\bm{u})\|_* 
    \geq \|L(\bm{\bar{w}})\|_* + \langle \bm{U}_1 \bm{V}_1^* + \bm{H}, L(\bm{u}) \rangle \\
    &= \|L(\bm{\bar{w}})\|_* + \langle \bm{U}_1 \bm{V}_1^* + \bm{H}, \mathcal{P}(L(\bm{u})) + \mathcal{P}^\perp(L(\bm{u})) \rangle.
\end{aligned}
\end{equation}
We expand the inner product:
\begin{align*}
    &\langle \bm{U}_1 \bm{V}_1^*, \mathcal{P}(L(\bm{u})) \rangle 
    + \langle \bm{U}_1 \bm{V}_1^*, \mathcal{P}^\perp(L(\bm{u})) \rangle \\
    &\qquad\qquad + \langle \bm{H}, \mathcal{P}(L(\bm{u})) \rangle 
    + \langle \bm{H}, \mathcal{P}^\perp(L(\bm{u})) \rangle.
\end{align*}
The first term satisfies
\[
|\langle \bm{U}_1 \bm{V}_1^*, \mathcal{P}(L(\bm{u})) \rangle| \leq \|\bm{U}_1 \bm{V}_1^*\|_2 \|\mathcal{P}(L(\bm{u}))\|_* = \|\mathcal{P}(L(\bm{u}))\|_*.
\]
The second and third terms vanish due to orthogonality, and the last term is
\[
\operatorname{Tr}((\bm{U}_2 \bm{V}_2^*)^* \bm{U}_2 \bm{\Sigma}_2 \bm{V}_2^*) = \operatorname{Tr}(\bm{V}_2 \bm{\Sigma}_2 \bm{V}_2^*) = \|\mathcal{P}^\perp(L(\bm{u}))\|_*.
\]
Substituting these four terms into \eqref{eq:Lw_Lu_geq_Lw_innprod_UVH_Lu} yields
\[
\|L(\bm{\bar{w}}) + L(\bm{u})\|_* \geq \|L(\bm{\bar{w}})\|_* - \|\mathcal{P}(L(\bm{u}))\|_* + \|\mathcal{P}^\perp(L(\bm{u}))\|_*.
\]
Now substitute this lower bound into \eqref{eq:Lwdot_Lu_leq_Lwdot_alpha} to obtain
\begin{align*}
    &\|L(\bm{\bar{w}})\|_* - \|\mathcal{P}(L(\bm{u}))\|_* + \|\mathcal{P}^\perp(L(\bm{u}))\|_* \\
    &\quad \leq \|L(\bm{\bar{w}})\|_* + \alpha \|L(\bm{u})\|_* \\
    &\quad \leq \|L(\bm{\bar{w}})\|_* + \alpha \|\mathcal{P}(L(\bm{u}))\|_* + \alpha \|\mathcal{P}^\perp(L(\bm{u}))\|_*.
\end{align*}
The cancellation of $\|L(\bm{\bar{w}})\|_*$ and the rearranging of terms gives
\begin{equation}
\label{eq:P_perp_Lu_leq_P_Lu}
(1 - \alpha) \|\mathcal{P}^\perp(L(\bm{u}))\|_* \leq (1 + \alpha) \|\mathcal{P}(L(\bm{u}))\|_*.
\end{equation}
In addition, we have
\begin{equation*}
\begin{aligned}
    \norm{L(\bm{u})}_* &\leq \norm{\mathcal{P}(L(\bm{u}))}_* + \norm{\mathcal{P}^\perp(L(\bm{u}))}_* \\
    &\leq \frac{2}{1 - \alpha} \norm{\mathcal{P}(L(\bm{u}))}_*,
\end{aligned}
\end{equation*}
where we used \eqref{eq:P_perp_Lu_leq_P_Lu} to derive the last inequality. By the relationship between the nuclear norm and the Frobenius norm, given by $\norm{\bm{X}}_* \leq \sqrt{\operatorname{Rank}(\bm{X})} \norm{\bm{X}}_F$, we obtain
\begin{equation}
\label{eq:L_nuc_leq_kappa_F}
\begin{aligned}
 &   \norm{L(\bm{u})}_* \leq \frac{2}{1 - \alpha} \norm{\mathcal{P}(L(\bm{u}))}_* \\ & \leq \frac{2\sqrt{2\kappa}}{1 - \alpha} \norm{\mathcal{P}(L(\bm{u}))}_F 
    \leq \frac{2\sqrt{2\kappa}}{1 - \alpha} \norm{L(\bm{u})}_F,
\end{aligned}
\end{equation}
where we used that $\operatorname{rank}(\mathcal{P}(X)) \le 2\kappa$ for $X\in\mathbb{C}^{M\times M}$ because $\mathcal{P}(X)=U_1U_1^{}X+XV_1V_1^{}-U_1U_1^{}XV_1V_1^{}$ with $\operatorname{rank}(U_1)=\operatorname{rank}(V_1)=\kappa$, and that $\|\mathcal{P}(\bm{X})\|_F \le \|\bm{X}\|_F$ for any orthogonal projector $\mathcal{P}$.

It remains to bound $\norm{L(\bm{u})}_F$ in terms of $\norm{\bm{u}}_2$. This is established by the following lemma, which relates the Frobenius norm of the Loewner operator to the $\ell_2$ norm of its argument via a numerical integration argument:

\begin{lemma}[Proof in Appendix~\ref{sec:useful}]
\label{lem:L_F_to_u_2_norm_bound}
    Consider a stable LTI system with transfer function
    \(G \in \mathcal{H}_\infty(\rho,K_G)\). Let \(\bm{z} \in \mathbb{C}^M\) satisfy Assumption~\ref{ass:z_margin}, and let \(\bm{w}\) be the frequency response of \(G\) at these points. If \(h_{\max}\) and \(\delta_m\) are sufficiently small (The explicit smallness conditions are given in the proof in the Appendix~\ref{sec:useful}), then
    \begin{equation}
    \label{eq:L_F_to_u_2_norm_bound}
        \frac{\norm{L(\bm{w})}_F}{\norm{\bm{w}}_2}
        \leq
        3\sqrt{1 + \frac{2K_G}{\rho - 1}}
        \sqrt{\frac{-h_{\max}\ln\sin\delta_m}{\pi h_{\min}^2}} .
    \end{equation}
\end{lemma}
Since $\bm u\in\mathcal T_{\bm{\bar w}}^{\rm st}\subseteq \mathcal W_M(\rho,2K)$, there exists \(G_u\in\mathcal H_\infty(\rho,2K)\) such that \(G_u(z_r)=u_r\) for all \(r\). Hence Lemma~\ref{lem:L_F_to_u_2_norm_bound} applies to $\bm u$ with \(K_G=2K\).
Applying Lemma~\ref{lem:L_F_to_u_2_norm_bound} to bound $\norm{L(\bm{u})}_F$ in \eqref{eq:L_nuc_leq_kappa_F}, we have that for sufficiently small $h_{\max}$ and $\delta_m$
\begin{equation*}
    \norm{L(\bm{u})}_* \leq \frac{6\sqrt{2\kappa}}{1-\alpha}\sqrt{1+\frac{4K}{\rho-1}}
    \sqrt{\frac{-h_{\max}\ln\sin\delta_m}{\pi h_{\min}^2}}\norm{\bm{u}}_2.
\end{equation*}
Rearranging the terms, we obtain
\begin{equation*}
    \frac{\norm{\bm{u}}_2}{\norm{L(\bm{u})}_*} \geq
    \frac{(1-\alpha)\,h_{\min}\sqrt{\pi}}
    {6\sqrt{1+\frac{4K}{\rho-1}}\sqrt{-2\kappa h_{\max}\ln\sin\delta_m}}.
\end{equation*}
Taking the infimum over all nonzero $\bm u\in\mathcal T_{\bm{\bar w}}^{\rm st}$ gives the following lower bound for the stable minimum gain:
\begin{equation}
\label{eq:phi_bound}
    \phi \geq \frac{(1-\alpha)\,h_{\min}\sqrt{\pi}}
    {6\sqrt{1+\frac{4K}{\rho-1}}\sqrt{-2\kappa h_{\max}\ln\sin\delta_m}}.
\end{equation}
Note that the inequality \eqref{eq:L_nuc_leq_kappa_F} can be replaced by the trivial bound $\|L(\bm{u})\|_* \leq \sqrt{M} \|L(\bm{u})\|_F$, and this trivial bound could be sharper. Thus, a better approach is to use the inequality
$
\|L(\bm{u})\|_* \leq \min \left\{\frac{2\sqrt{2\kappa}}{1-\alpha}, \sqrt{M}\right\} \|L(\bm{u})\|_F.
$
Nonetheless, we choose only the $\kappa$-dependent bound since we are interested in the dependence of the sample complexity bound in the regime where $M$ is large, while $\kappa$ is small.

\subsection{Bounding the Noise-Dependent Term}
\label{sec:noise_term}

Finally, to complete the derivation of the sample complexity's bound, we establish an upper bound on the maximum error associated with the inverse adjoint of the Loewner operator of the noise: $\| L^{-1}(\bm{v}_s) \|_2$. To this end, let Assumptions \ref{ass:noise} and \ref{ass:z_margin} hold. Leverage the explicit formula for the inverse adjoint Loewner operator from \eqref{eq:loewner_inv_conj} in Appendix~\ref{sec:adj_inv_loewner}:
\begin{equation}
\label{eq:average_error_expansion}
\begin{aligned}
    &L^{-*}(\bm{v}_s) = L^{-*}(\begin{bmatrix}
        \bm{a}_s \\ \bm{b}_s
    \end{bmatrix}) \\
    &= \sum_{k=1}^M \left( \bm{e}_k^\top \Ein^{\dagger} \bm{a}_s \bm{E}_k \right) + \sum_{k=1}^M \left( \bm{e}_k^\top \Fin^{-1} \bm{b}_s \bm{F}_k \right).
\end{aligned}
\end{equation}
To upper bound $\| L^{-*}(\bm{v}_s)\|_2$, we derive lower bounds on the smallest singular values of the matrices $\Ein$ and $\Fin$\footnote{Since $\Ein$ and $\Fin$ are positive semidefinite matrices, their eigenvalues and singular values are identical. 
} These matrices exhibit a specific structure, allowing us to establish a lower bound on their minimum nonzero singular values as follows:

For the matrix $\Ein$, examining \eqref{eq:E_in} reveals that it is the Laplacian matrix of a graph. This observation aligns with the fact that its null space is one-dimensional, with $\bm{1}_M$ being the associated eigenvector. Furthermore, each entry of $\Ein$ has an absolute value of at least $0.5$. Consequently, by applying Lemma~\ref{lem:alg_connc_lower_bound}, we establish that the algebraic connectivity, defined as the smallest nonzero eigenvalue, of $\Ein$ (in this case: $\lambda_{M-1}(\Ein)$) is at least $M/2$. As a result, we obtain the bound
\begin{equation}
\label{eq:E_in_dagger_bound}
    \| \Ein^\dagger \|_2 \leq \frac{2}{M}.
\end{equation}

Consider a term in the first summation in the RHS of \eqref{eq:average_error_expansion}: $e_k \Ein^\dagger\bm{a}_s \bm{E}_k$. We may bound this term as follows
\begin{equation}
\label{eq:E_in_term_bound}
\begin{aligned}
    &\| e_k^\top \Ein^\dagger \bm{a}_s \bm{E}_k \|_2 \leq | e_k^\top \Ein^\dagger \bm{a}_s | \| \bm{E}_k \|_2 \\
    &\leq \|\Ein^\dagger\|_2 \|\bm{a}_s\|_2 \|\bm{E}_k\|_2 \leq \frac{2\bar{\eta}}{M} \| \bm{E}_k \|_F
\end{aligned}
\end{equation}
in which we used \eqref{eq:E_in_dagger_bound}, Assumption~\ref{ass:noise}, and the relation $\| \cdot \|_2 \leq \| \cdot \|_F$ to conclude the last inequality.

For the matrix $\Fin$, examining \eqref{eq:F_in} reveals that we can decompose it as
\begin{equation*}
    \Fin = \Gin + \Din
\end{equation*}
in which $(\Gin)_{rs} = \frac{2}{|\bar{z}_r - z_s|^2}$ and $\Din$ is a digonal matrix such that $ (\Din)_{rr} = \sum_{k} \left( \frac{2}{| \bar{z}_r - z_k |^2} \right)$. Lemma~\ref{lem:pos_semi_def_kernel} which follows from Theorem~\ref{thm:inverse_multiquadratics} proves that $\Gin$ is positive semi-definite. Therefore, 
\begin{equation*}
\begin{aligned}
    \lambda_{\min}(\Fin) &\geq \lambda_{\min}(\Din) = \min_r D_{rr} \\
    &= \min_r \sum_{k} \left( \frac{2}{| \bar{z}_r - z_k |^2} \right) \geq \sum_{k} \frac{2}{4} = \frac{M}{2},
\end{aligned}
\end{equation*}
where we used $| \bar{z}_r - z_k | \leq 2$ to derive the second inequality. Consequently, $\| \Fin^{-1} \|_2 \leq \frac{2}{M}$. Similar to \eqref{eq:E_in_term_bound}, we may bound a term in the second summation in the RHS of \eqref{eq:average_error_expansion}, $\bm{e}_k^\top \Fin^{-1} \bm{b}_s \bm{F}_k$, as follows
\begin{equation}
\label{eq:F_in_term_bound}
    \| \bm{e}_k^\top \Fin^{-1} \bm{b}_s \bm{F}_k \|_2 \leq \frac{2 \bar{\eta}}{M} \| \bm{F}_k \|_F.
\end{equation}

We aim to apply the concentration inequality in Theorem~\ref{thm:matrix_hoeffding} to $\sum_{s=1}^N L^{-*}(\bm{v}_s)$. In this case, we may choose $\bm{A}_k$ in Theorem~\ref{thm:matrix_hoeffding} to be $\frac{2\bar{\eta}}{M}\|\bm{E}_k\|_F \bm{I}_M$ and $\frac{2\bar{\eta}}{M}\|\bm{F}_k\|_F \bm{I}_M$ for the first and second summations in \eqref{eq:average_error_expansion} respectively. Thus, the variance $\sigma^2$ in Theorem~\ref{thm:matrix_hoeffding} is:

\begin{equation*}
\begin{aligned}
    \sigma^2 &= \left\| \sum_{s=1}^N \left( \sum_{k=1}^M \frac{4\bar{\eta}^2}{M^2} \|\bm{E}_k\|_F^2 \bm{I}_M + \sum_{k=1}^M \frac{4\bar{\eta}^2}{M^2} \|\bm{F}_k\|_F^2 \bm{I}_M \right) \right\|_2 \\
    &= \frac{4 N \bar{\eta}^2}{M^2} \sum_{k=1}^M \left( \| \bm{E}_k \|_F^2 + \| \bm{F}_k \|_F^2 \right) \\
    &= \frac{4 N \bar{\eta}^2}{M^2} \sum_{k=1}^M \left( \sum_{s\neq k} \frac{2}{|\bar{z}_k - z_s|^2} + \right. \\
    & \qquad\qquad\qquad \left. \sum_{s \neq k} \frac{2}{| \bar{z}_k - z_s |^2} + \frac{4}{| \bar{z}_k - z_k |^2} \right) \\
    &= \frac{16 N \bar{\eta}^2}{M^2} \sum_{k=1}^M \sum_{s=1}^M \frac{1}{| \bar{z}_k - z_s |^2}.
\end{aligned}
\end{equation*}
To bound $\sigma$, we need to control the double sum $\sum_{k=1}^M \sum_{s=1}^M \frac{1}{|\bar{z}_k - z_s|^2}$. This is established by the following lemma, which bounds the sum by a trigonometric integral:

\begin{lemma}[Proof in Appendix~\ref{sec:useful}]
\label{lem:sum_z_bound_int}
    Let $\bm{z} \in \mathbb{C}^M$ satisfy Assumption~\ref{ass:z_margin}. Then,
    \begin{equation}
    \label{eq:sum_1_overzsq_leq_lnsin_delta}
        \sum_{r=1}^M \sum_{s=1}^M \frac{1}{|\bar{z}_r - z_s|^2} \leq \frac{-2\ln\sin\delta_m}{h_{\min}^2}.
    \end{equation}
\end{lemma}

Consequently by the application of Lemma~\ref{lem:sum_z_bound_int}, $\sigma$ is bounded by:
\begin{equation}
\label{eq:delta_not_dep_M}
    \sigma \leq \frac{4\bar{\eta}\sqrt{-2N\ln\sin\delta_m}}{M h_{\min}}.
\end{equation}
Surprisingly, for uniform grids with $h_{\min} = (\pi-2\delta_m)/M$, $\sigma$ is independent of $M$. Finally, by choosing $t=2\sqrt{2}\sigma\sqrt{\ln(\frac{M}{\delta})}$ in \eqref{eq:exp_var_bound_conc_measure}, we obtain
\begin{equation}
\label{eq:mu_sigma_bound}
\begin{aligned}
    &\mathbb{P} \left( \| \sum_{s=1}^N L^{-*}(\bm{v}_s) \|_2 \geq  \right. \\
    &\qquad\qquad \left. \frac{16\bar{\eta}\sqrt{-N\ln\sin\delta_m\,\ln(M/\delta)}}{M h_{\min}} \right) \leq \delta.
\end{aligned}
\end{equation}

\subsection{Sampled-Frequency Estimation Error}
\label{sec:sampled_frequency_bound}

Having established the minimum-gain bound \eqref{eq:phi_bound} and the noise
concentration bound \eqref{eq:mu_sigma_bound}, we are now ready to prove the main result.

\begin{proof}[Proof of Theorem~\ref{thm:sample_complexity_finite_sample}]
By substituting $\phi$ from \eqref{eq:phi_bound} and the bound from \eqref{eq:mu_sigma_bound} into \eqref{eq:finite_sample_general_bound}, we derive the following sample complexity bound as a consequence of Lemma~\ref{lem:finit_sample_bound}:

\begin{equation}
\label{eq:finite_sample_bound_with_alpha}
    \begin{aligned}
        & \| \bm{\hat{w}} - \bm{\bar{w}} \|_\infty \leq \\
        & \frac{12\sqrt{2\kappa}\sqrt{1+\frac{4K}{\rho-1}}
        \sqrt{-h_{\max}\ln\sin\delta_m}}
        {(1-\alpha)h_{\min}\sqrt{\pi}}
        \left( \frac{\tau}{MN} \right. \\
        & \left. + \frac{16\bar{\eta}\sqrt{-\ln\sin\delta_m\,\ln(M/\delta)}}
        {M h_{\min}\sqrt{N}} \right).
    \end{aligned}
\end{equation}

Next, using the bound in \eqref{eq:mu_sigma_bound} we can upper bound $\alpha$ given in \eqref{eq:alpha} as:
\begin{equation}
    \label{eq:alpha_upper_bound}
    \alpha \leq \frac{16\bar{\eta}\sqrt{-N\ln\sin\delta_m\,\ln(M/\delta)}}{\tau\,h_{\min}}
\end{equation}
The substitution of this bound into \eqref{eq:finite_sample_bound_with_alpha} removes the dependency on $\alpha$ and yields an expression of the form $D\frac{A\tau + B}{1 - \frac{C}{\tau}}$, where $A$, $B$, $C$, and $D$ are independent of $\tau$. This bound attains its minimum at $\tau^\star = C + \sqrt{C^2 + \frac{C B}{A}}$ resulting in \eqref{eq:tau_optimal}.
Interestingly, substitution of this optimal $\tau^*$ into \eqref{eq:alpha_upper_bound} yields the bound $\alpha \leq \sqrt{2}-1 < 1$. Finally, substitution of $\tau=\tau^*$ and $\alpha=\sqrt{2} - 1$ into \eqref{eq:finite_sample_bound_with_alpha} yields the final bound in \eqref{eq:finite_sample_bound} completing the proof.
\end{proof}

\section{Numerical Example}
\label{sec:numerical_example}

We present numerical experiments to validate our main results. For the SISO experiments in this section, we use the following system from \cite{tsiamis2024finitesamplefrequencydomain} as the ground-truth model:

\begin{equation}
\label{eq:true_system}
    G(z) = \frac{0.12 z + 0.18 z^{2}}{1 - 1.4 z^{1} + 1.443 z^{2} - 1.123 z^{3} + 0.7729 z^{4}}.
\end{equation}

\subsection{Estimating a low order frequency response from noisy data}
\label{sec:low_order_identification}

In this experiment, we demonstrate that \eqref{eq:LNNM}, with an appropriately chosen  regularization parameter, effectively promotes low-order representations. We use $M=32$  frequency points logarithmically spaced over $[0,\pi]$, allocating more samples to the  low-frequency range where control-relevant dynamics are most significant. The experiment  parameters are set as follows: $N=30$ measurements per frequency, noise bound  $\bar{\eta}=2$, regularization parameter $\tau=25$, and prior parameters $\rho=1.01$ and $K=50$. The proposed method is benchmarked against two baselines: the classical  averaging estimator and a smoothed variant obtained by applying a Gaussian moving-average  window of width~5 to the averaged frequency response estimate.

\begin{figure}
    \centering
    \includegraphics[width=0.8\linewidth]{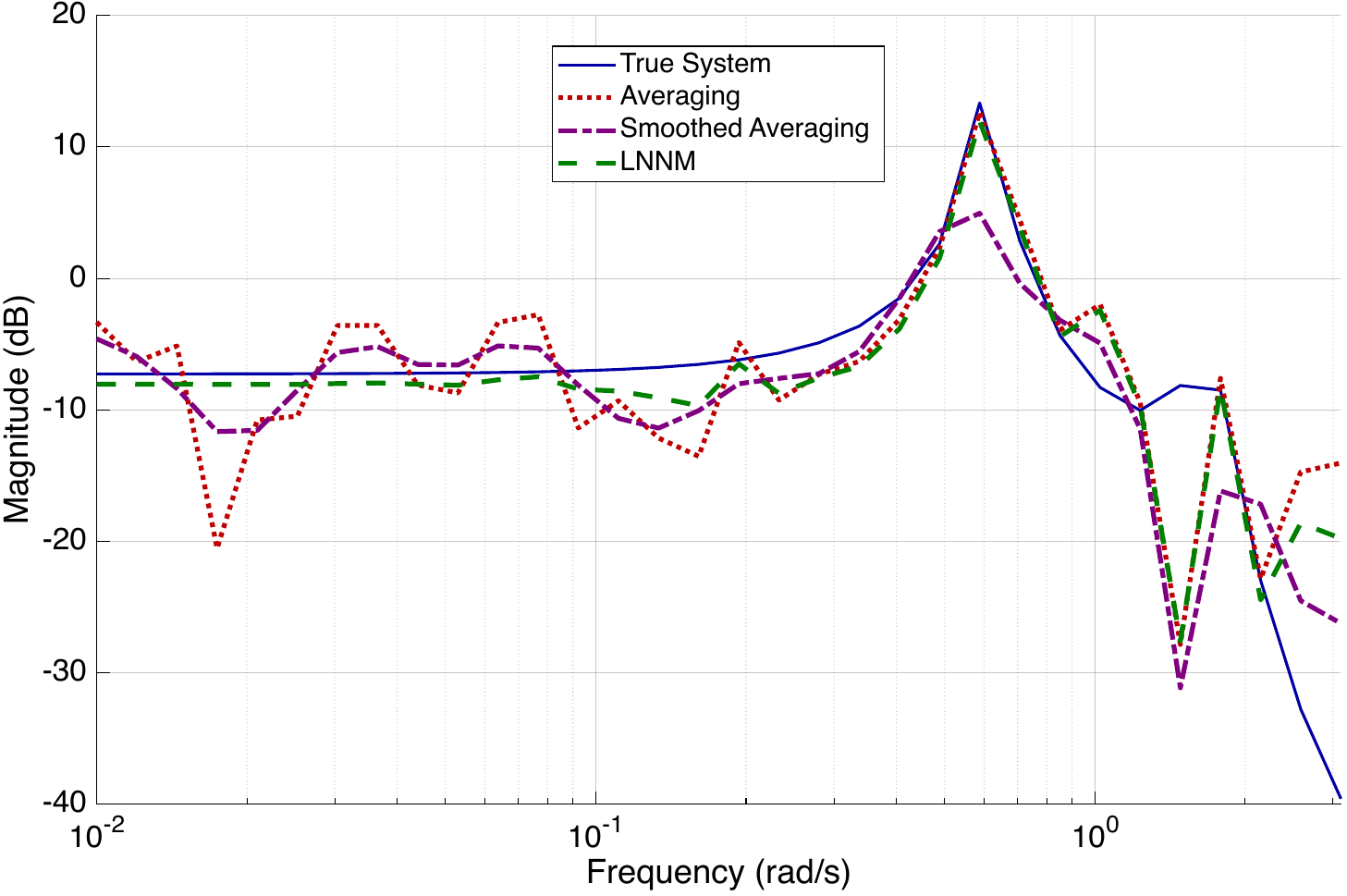}
    \caption{\emph{Bode Diagrams:} The solid blue line represents the frequency response of the \textcolor{myblue}{ground truth system}. The frequency responses obtained by \textcolor{myred}{Averaging} (red dotted), \textcolor{mypurple}{Smoothed Averaging} (purple dash-dot), and \textcolor{mygreen}{LNNM} (green dashed) are compared to the baseline. LNNM exhibits less fluctuation at low frequencies, resulting in a response closer to the ground truth. Smoothed Averaging reduces high-frequency noise but introduces bias near the resonance peak.}
    \label{fig:bode_comparison}
\end{figure}

\begin{figure}
    \centering
    \includegraphics[width=0.8\linewidth]{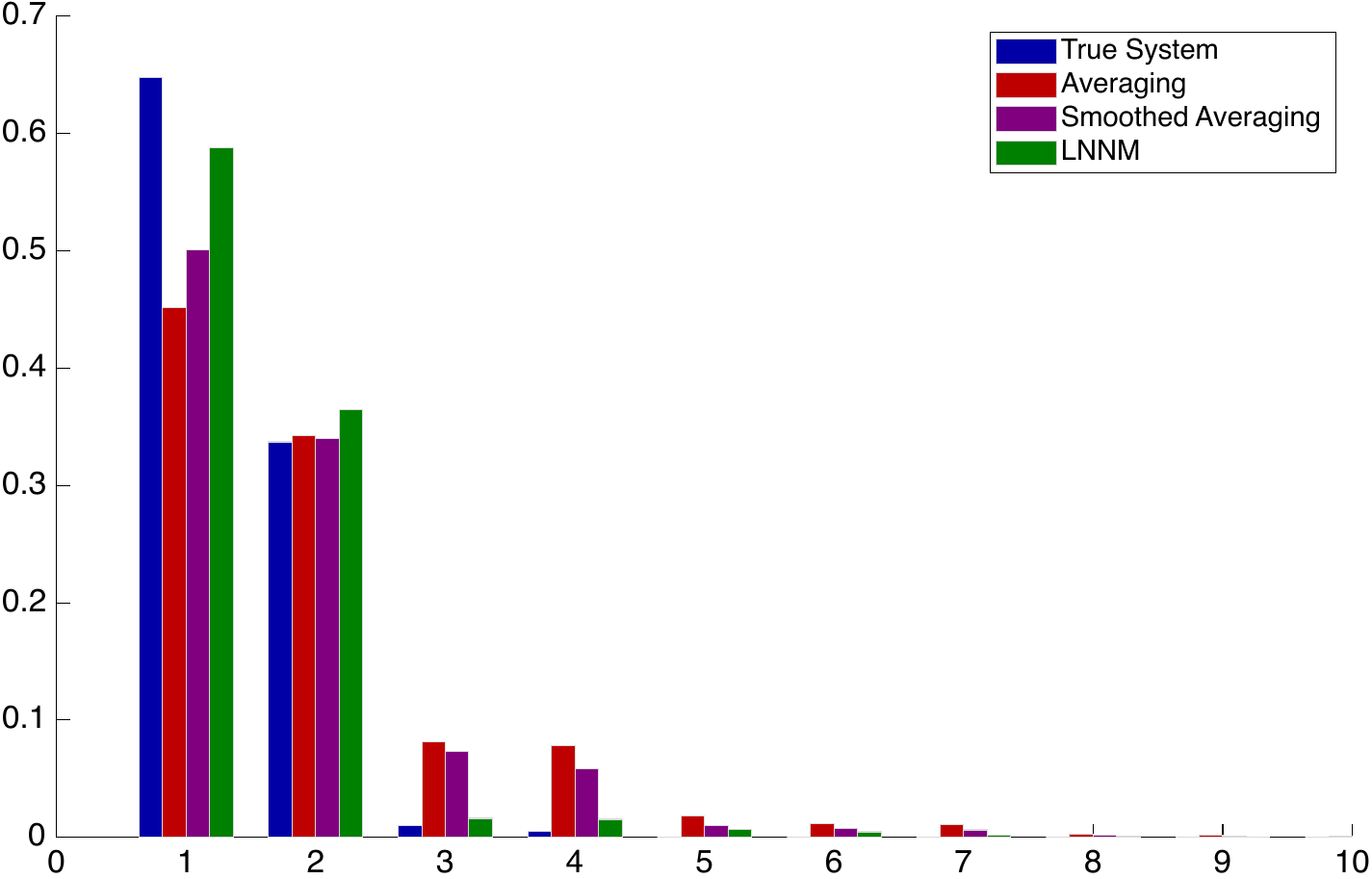}
    \caption{\emph{Singular Values of the Loewner Matrix:} The singular values are normalized to have unit sum, representing the relative contribution of each mode. For \textcolor{mygreen}{LNNM}, they are concentrated in the leading modes and decay rapidly, closely matching the low-order \textcolor{myblue}{ground truth system}. In contrast, both \textcolor{myred}{Averaging} and \textcolor{mypurple}{Smoothed Averaging} produce spread-out distributions, indicating a larger effective order. This shows that frequency-domain smoothing reduces noise but does not promote low order structure, unlike LNNM. As the McMillan degree of the identified system equals the numerical rank of its Loewner matrix; hence, the rapid singular value decay for LNNM indicates a lower-order realization compared to Averaging.}
    \label{fig:svd_comparison}
\end{figure}

Figure~\ref{fig:bode_comparison} compares the frequency response estimates produced by averaging, smoothed averaging, and LNNM against the true system \eqref{eq:true_system}. Figure~\ref{fig:svd_comparison} further illustrates the advantage of the proposed approach by comparing the singular values of the Loewner matrices across all three methods. Since the rank of the Loewner matrix equals the order of the lowest-order rational interpolant of the frequency-domain data, a faster singular value decay directly indicates a lower-order realization---a behavior markedly more pronounced under LNNM.

\subsection{Numerical verification of Sample Complexity}
\label{sec:sample_complexity_verification}

\begin{figure}
    \centering
    \includegraphics[width=0.8\linewidth]{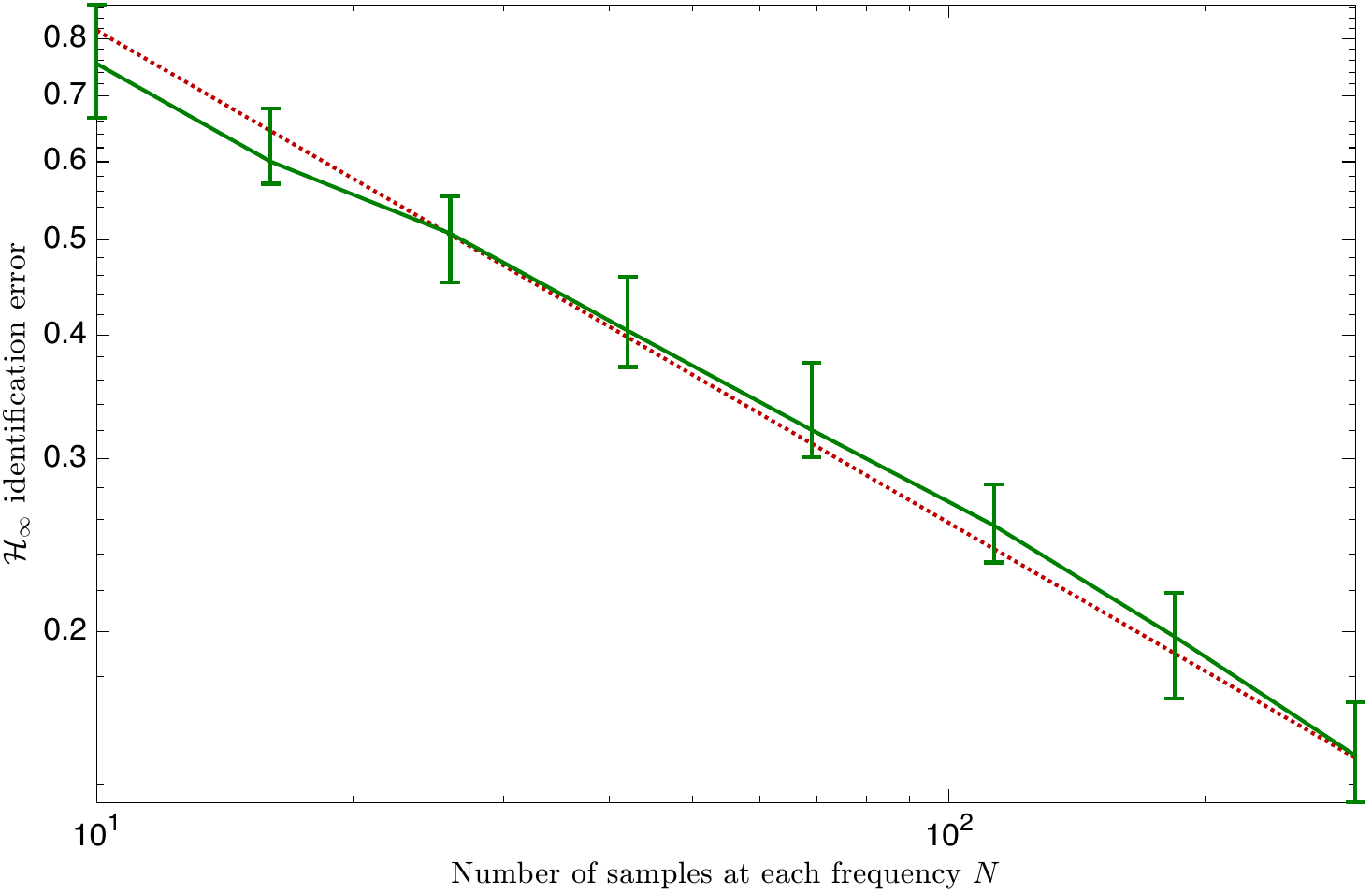}
    \caption{\emph{Sampled-Frequency Identification Error as a Function of $N$} The results are presented in a logarithmic scale, where the red dotted line represents a reference slope of $0.5$, confirming the $\mathcal{O}(N^{-\frac{1}{2}})$ scaling in sample complexity.}
    \label{fig:N_dep}
\end{figure}

\begin{figure}
    \centering
    \includegraphics[width=0.8\linewidth]{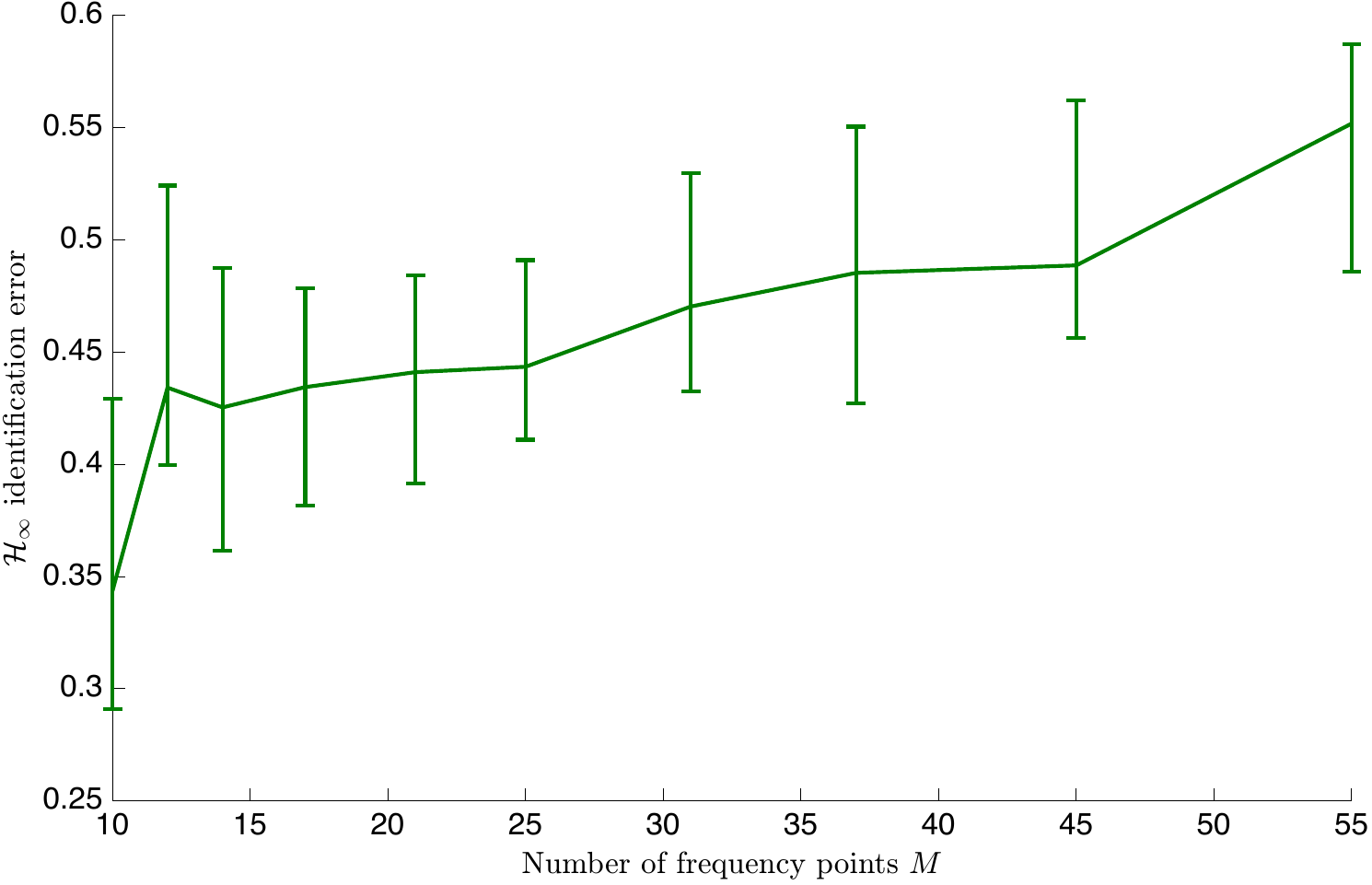}
    \caption{\emph{Sampled-Frequency Identification Error as a Function of $M$.} The identification error grows slowly with $M$. The theoretical rate $\mathcal{O}((M\ln M)^{1/2})$ arises from a worst-case bound on the Loewner operator gain; the mild empirical growth suggests this bound is conservative and could potentially be tightened through a more refined analysis.}
    \label{fig:M_dep}
\end{figure}

\begin{figure}
    \centering
    \includegraphics[width=0.8\linewidth]{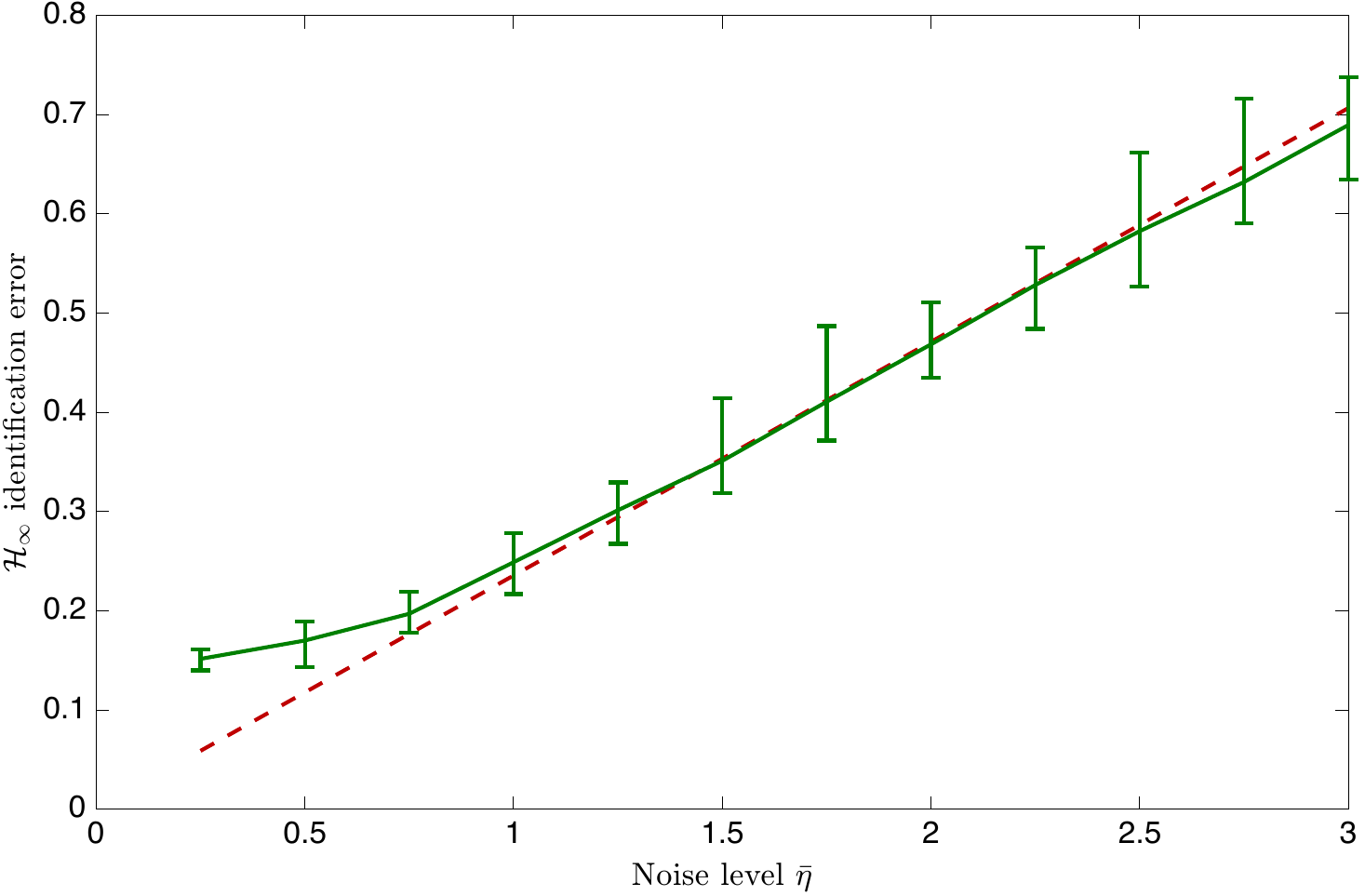}
    \caption{\emph{Sampled-Frequency Identification Error as a Function of $\bar\eta$} illustrating the linear dependency on the noise level, $\mathcal{O}(\bar{\eta})$.}
    \label{fig:eta_dep}
\end{figure}

In this section, we examine the dependency of the sample complexity bound in \eqref{eq:finite_sample_bound} with respect to $N$, $M$, and $\bar{\eta}$. Monte Carlo simulations comprising 50 independent experiments were conducted to analyze the dependency of the sampled-frequency estimation error. Figures~\ref{fig:N_dep}, \ref{fig:M_dep}, and \ref{fig:eta_dep} present plots of the sampled-frequency $\mathcal{H}_\infty$ identification error versus $N$, $M$, and $\bar{\eta}$, respectively.

The experimental parameters are chosen as in Section~\ref{sec:low_order_identification}, with $N=30$, $M=32$, $\bar{\eta}=2$, $\tau=25$, $\rho=1.01$, and $K=50$. The frequency points are uniformly spaced over $[\delta_m,\pi-\delta_m]$ with $\delta_m=0.01$, to match the setting of the finite-sample analysis. When one of the parameters $N$, $M$, or $\bar{\eta}$ is varied on the horizontal axis, the remaining parameters are kept fixed at the above values. With these parameters, the finite-sample bound in Theorem~\ref{thm:sample_complexity_finite_sample} evaluates to approximately \(10^6\), while the empirical median \(\mathcal{H}_\infty\) error is approximately \(0.3\). This confirms that the numerical prefactor is conservative, as expected for a data-independent non-asymptotic guarantee. The main role of the theorem is therefore to characterize parameter dependence.

\begin{figure}
    \centering
    \includegraphics[width=0.8\linewidth]{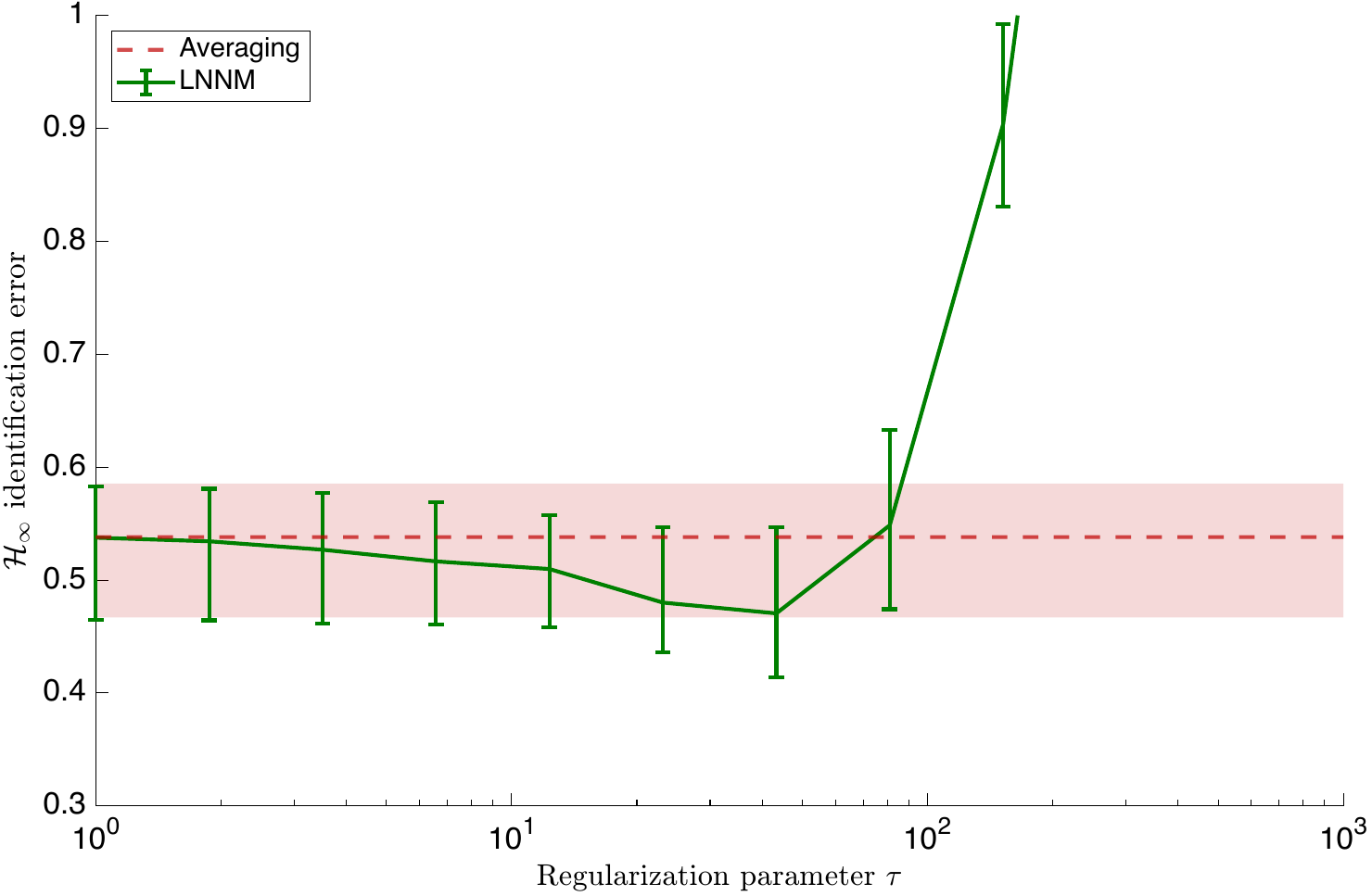}
    \caption{\emph{$\mathcal{H}_\infty$ Identification Error as a Function of $\tau$:}
    The median error of \textcolor{mygreen}{LNNM} remains below
    \textcolor{myred}{Averaging} over two orders of magnitude of $\tau$.
    The smoothed-averaging curve is omitted from this plot because it lies
    substantially above both methods; in this example, the smoothing window
    blurs the resonance peak visible in Figure~\ref{fig:bode_comparison},
    introducing bias that dominates the variance reduction.}
    \label{fig:tau_dep}
\end{figure}

Figure~\ref{fig:tau_dep} examines the sensitivity of the $\mathcal{H}_\infty$ identification error to the choice of $\tau$. The median error of LNNM remains below that of Averaging over two orders of magnitude of $\tau$, with approximately $10\%$ lower error in the optimal range $\tau \in [30, 60]$. For $\tau > 100$, over-regularization dominates and the error increases. This confirms that while cross-validation can be used to tune $\tau$ optimally, the method is robust to the exact choice of $\tau$ over a wide range.

\subsection{Multi-System Comparison}
\label{sec:multi_system}

\begin{figure}
    \centering
    \includegraphics[width=\linewidth]{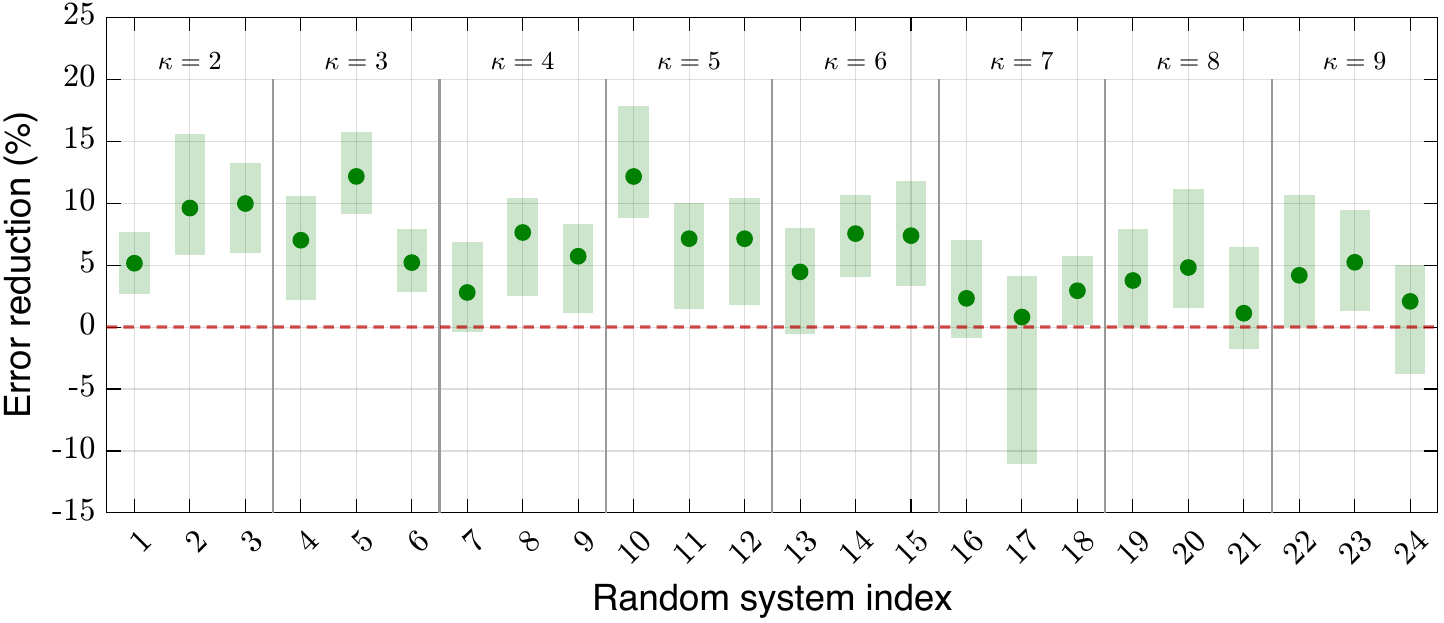}
    \caption{\emph{LNNM Error Reduction Across 24 Random Systems:} Median (circle) and interquartile range (shaded) of the percentage reduction in sampled-frequency $\mathcal{H}_\infty$ identification error of LNNM relative to Averaging, grouped by McMillan degree $\kappa$.}
    \label{fig:multi_system}
\end{figure}

To assess generality, we apply LNNM and averaging to 24 randomly generated stable SISO systems spanning McMillan degrees $\kappa \in \{2,3,\ldots,9\}$ (three systems per degree), using fixed parameters $M=32$, $N=30$, $\bar{\eta}=2$, $\tau=25$, $\rho=1.01$, and $K=50$ across all systems. These hyperparameters are not cross-validated per system; the comparison is purely illustrative. For each system, 50 Monte Carlo trials are run and the reduction in identification error is calculated with the results summerize in Figure~\ref{fig:multi_system}. LNNM achieves positive error reduction for the majority of systems, and per-system tuning of $(\tau, \rho, K)$ would be expected to further improve these results.

\subsection{Realization Comparison}
\label{sec:realization_comparison}

A natural question is whether a better frequency-response estimate translates into a better realized state-space model. To examine this, we apply two standard realization methods to each of the three frequency-response estimates considered in Section~\ref{sec:low_order_identification}: Averaging, Smoothed Averaging, and LNNM. The same fixed system and parameters are used: $M=32$, $N=30$, $\bar{\eta}=2$, $\tau=25$, $\rho=1.01$, and $K=50$. In all cases we select the model order $\kappa=4$, as suggested by the numerical rank of the Loewner matrix.

The two realization methods are:
\begin{enumerate}
\item \emph{Subspace identification:} via \texttt{n4sid} in System Identification Toolbox of MATLAB \cite{mckelvey1996subspace_freq, matlab_sysid_toolbox}.
\item \emph{Maximum Likelihood:} variance-weighted frequency-domain maximum-likelihood estimation via \texttt{mlfdi} from FdiTools \cite{pintelon2012system, fditools_github}, where the per-frequency noise variance of the frequency-response estimate is used as the weighting.
\end{enumerate}

\begin{figure}[t]
    \centering
    \includegraphics[width=\linewidth]{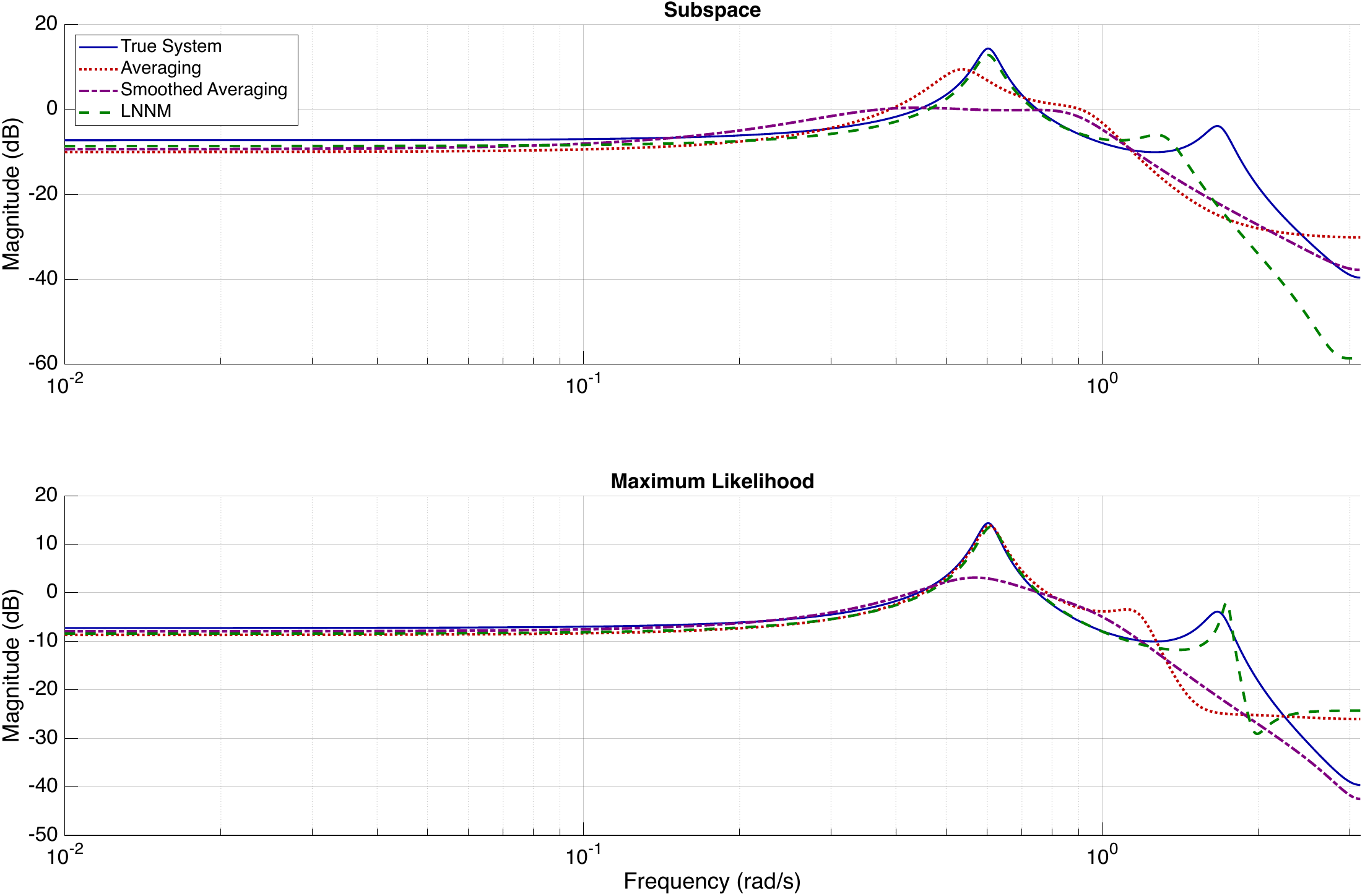}
    \caption{\emph{Realization Comparison:} Bode magnitude of realized models (order 4) for each combination of frequency estimator and realization method (Subspace and Maximum Likelihood). The \textcolor{myblue}{true system} (solid blue), \textcolor{myred}{Averaging} (red dotted), \textcolor{mypurple}{Smoothed Averaging} (purple dash-dot), and \textcolor{mygreen}{LNNM} (green dashed) are shown. LNNM-based realizations track the ground truth most closely under both methods.}
    \label{fig:realization_comparison}
\end{figure}

Figure~\ref{fig:realization_comparison} shows the Bode magnitude of the realized models. Across both realization methods, the model realized from the LNNM estimate more closely captures the resonance peak of the true system. Notably, Smoothed Averaging damps the resonance, which is a direct consequence of its smoothing bias, whereas LNNM preserves the peak through its low-order-promoting regularization.

\subsection{MIMO Identification}
\label{sec:mimo_identification}

\begin{figure}[t]
    \centering
    \includegraphics[width=\linewidth]{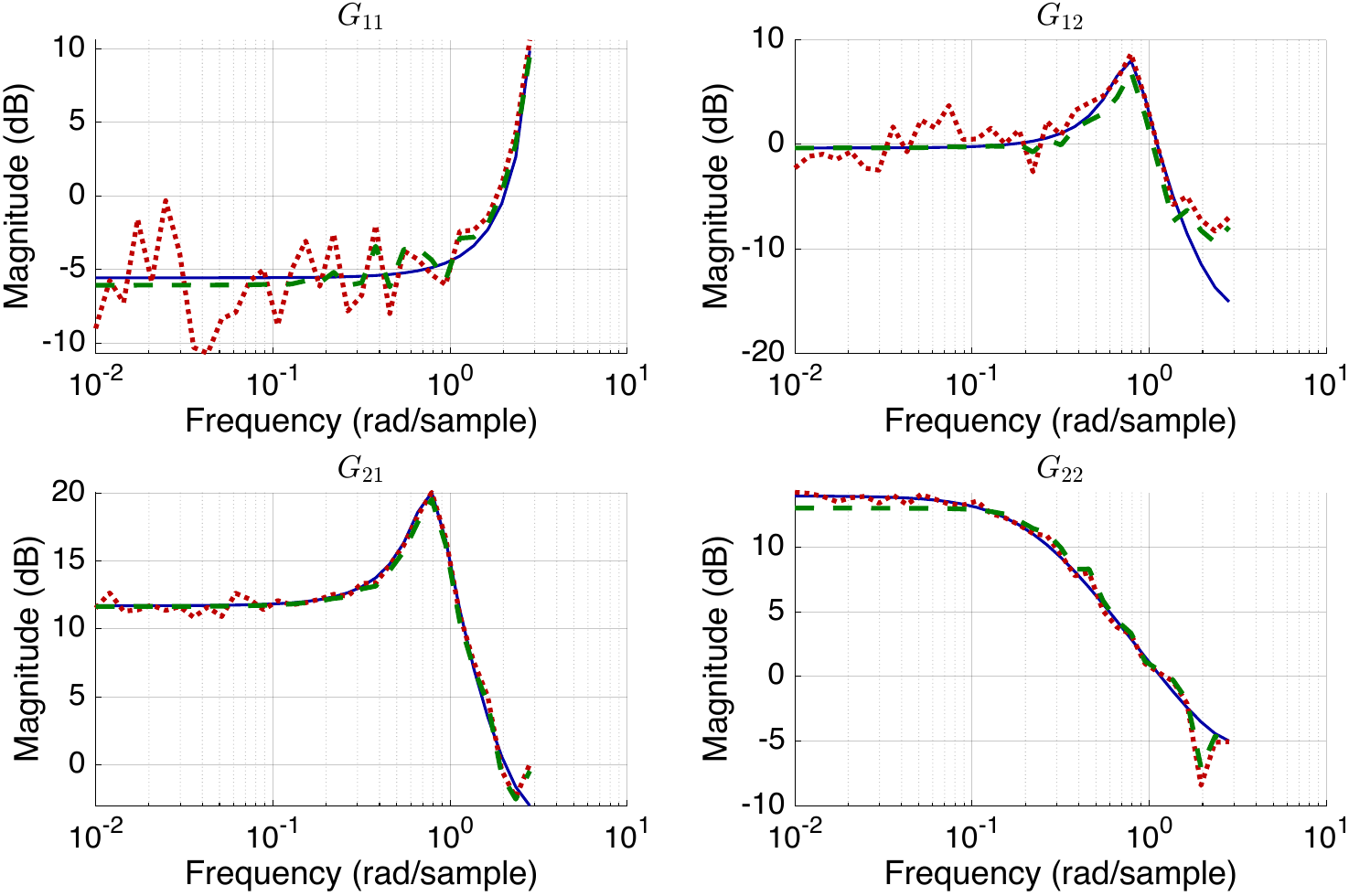}
    \caption{\emph{MIMO Bode Magnitude:} All four channels. \textcolor{myblue}{True system} (solid), \textcolor{myred}{Averaging} (dotted), \textcolor{mygreen}{LNNM} (dashed).}
    \label{fig:mimo_bode}
\end{figure}

\begin{figure}[t]
    \centering
    \includegraphics[width=0.9\linewidth]{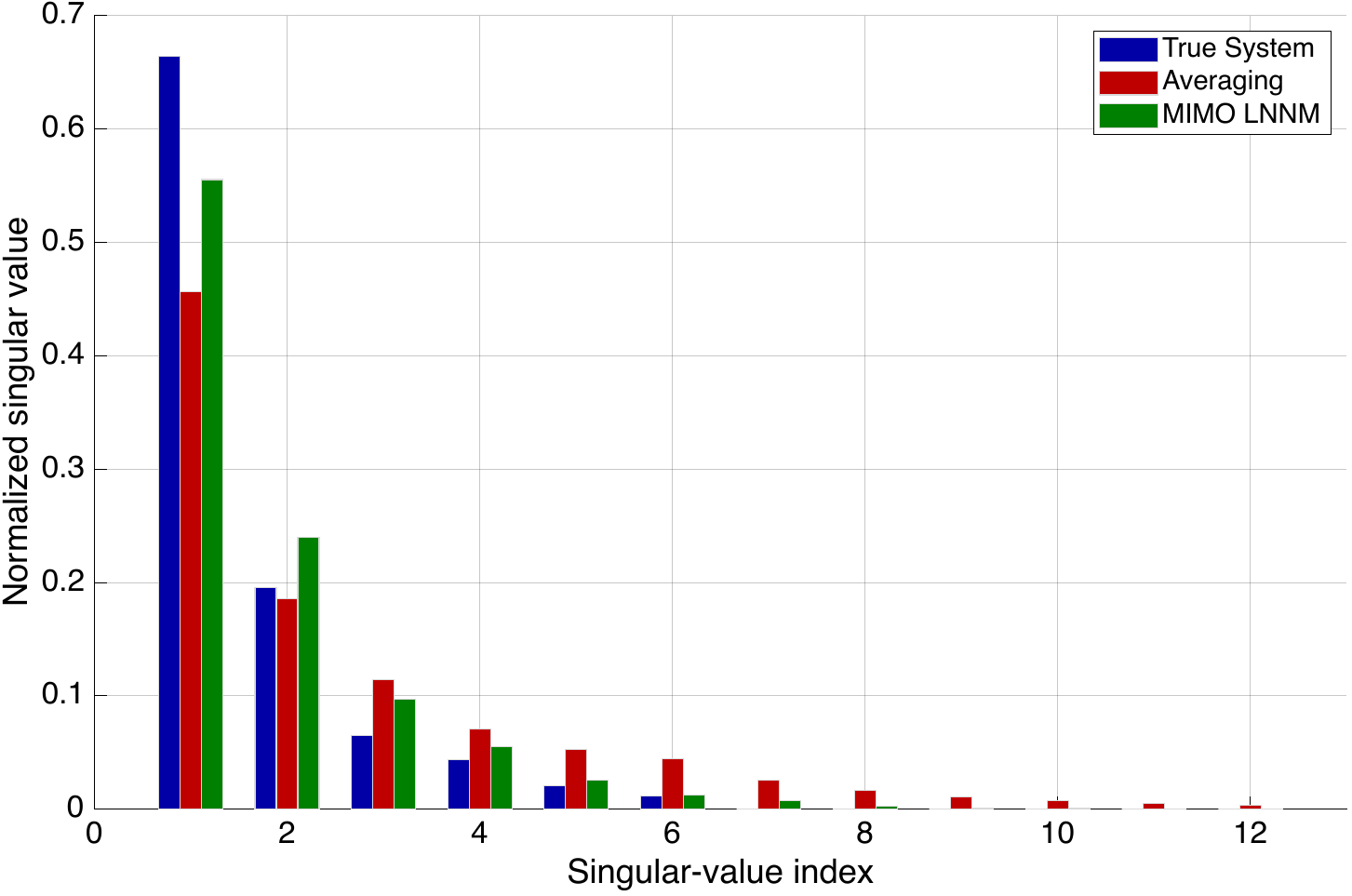}
    \caption{\emph{Block Loewner Singular Values (MIMO):} LNNM concentrates energy in the leading modes promoting low-order identification, unlike averaging.}
    \label{fig:mimo_svd}
\end{figure}

To demonstrate that \eqref{eq:LNNM} extends directly to MIMO systems, we apply it to the $2\times 2$ system
\begin{equation*}
\label{eq:mimo_system}
    \bar{G}(z) = \begin{bmatrix}
        \dfrac{z^{-1}}{1 + 0.9z^{-1}} & \dfrac{0.5z^{-2}}{1 - 1.2z^{-1} + 0.72z^{-2}} \\[10pt]
        \dfrac{2z^{-1}}{1 - 1.2z^{-1} + 0.72z^{-2}} & \dfrac{1}{1 - 0.8z^{-1}}
    \end{bmatrix}
\end{equation*}
of McMillan degree~$4$, using $M=32$, $N=30$, $\bar{\eta}=2$, $\tau=100$, $\rho = 1.05$, and $K = 50$. The block Loewner matrix $\mathcal{L}(\bm{W})\in\mathbb{C}^{Mn_y\times Mn_u}$ is formed entry-wise as $[\mathcal{L}(\bm{W})]_{ij} = (\overline{\bm{W}}_i - \bm{W}_j)/(\bar{z}_i - z_j)$, and the nuclear norm in \eqref{eq:LNNM} is taken over this block matrix. The sample complexity bound of Theorem~\ref{thm:sample_complexity_finite_sample} does not apply here. This experiment is purely demonstrative. Figure~\ref{fig:mimo_bode} shows that LNNM tracks the true frequency response more closely than averaging across all four channels. Figure~\ref{fig:mimo_svd} shows the normalized singular values of the block Loewner matrix: LNNM concentrates singular values in the leading modes, whereas averaging spreads them across a wider range of modes.

\section{Conclusions}
\label{sec:conclusion}

Research on low-order identification in the time domain, has led to non-asymptotic sample complexity bounds. However, a similar analysis for the frequency domain is not available. We propose an alternative, frequency-domain-based system identification method for learning low-order systems. Our main result provides an upper bound on the sample complexity of the identification process, shedding light on the different factors that affect it. Specifically, we establish that the sample complexity of the identification error scales as $\mathcal{O}(\sqrt{\frac{M \ln M}{N}})$. To the best of our knowledge, this is the first non-asymptotic bound for regularization-based frequency-domain algorithms. Several directions remain open. First, the present analysis could be conservative in its dependence on the number of frequency points \(M\). Second, although the optimization formulation extends naturally to MIMO systems through block Loewner matrices, extending the finite-sample theory to the MIMO setting is still open. Third, the quadratic data-fitting term in \eqref{eq:LNNM} could be replaced by other convex losses tailored to the measurement-noise model, such as weighted least-squares losses for frequency-dependent noise, robust losses for heavy-tailed errors, and quantile losses for asymmetric uncertainty \cite{kanakeri2025outlier, Khojaste2025quantile, shen2023computationally}.

\appendices

\section{Technical Backgrounds}
\label{sec:technical_bg}

\subsection{Concentration of Measure}
Concentration inequalities serve as powerful tools for establishing sample complexity bounds. Under reasonable assumptions regarding the distribution of noise, it is possible to analyze how certain random variables, such as the identification error, which depends on the noise distribution, concentrate around a specific value. The following theorem presents a matrix Hoeffding concentration inequality, which we later exploit in our analysis:

\begin{theorem}[Theorem 1.3 from \cite{tropp2012random_matrices}]
\label{thm:matrix_hoeffding}
    Consider a finite sequence $\left\{\bm{X}_s\right\}_{s=1}^N$ of independent, random, self-adjoint matrices with dimension $M$, and let $\left\{\bm{A}_s\right\}_{s=1}^N$ be a sequence of fixed self-adjoint matrices. Assume that each random matrix satisfies $\mathbb{E} (\bm{X}_s) = \bm{0}$ and $\bm{A}_s^2 - \bm{X}_s^2 \succeq 0$ almost surely. Then, for all $t \geq 0$,
    \begin{equation}
    \label{eq:exp_var_bound_conc_measure}
        \mathbb{P}\left\{ \lambda_{\max} \left(\sum_{s=1}^N \mathbf{X}_s\right) \geq t\right\} \leq M \mathrm{e}^{-t^2 / 8 \sigma^2} \quad
    \end{equation}
    where $\sigma^2=\left\|\sum_{s=1}^N \mathbf{A}_s^2\right\|_2$.
\end{theorem}


\subsection{Positive Definite Functions and Matrices}
As part of our analysis, we need to show that a certain type of matrix is positive definite. To do this, we use a special class of functions that help define such matrices:

\begin{definition}[Definition 6.15 from \cite{wendland2004scat_data_approx}]
\label{def:pos_def_func}
    We will call a univariate function $\phi: [0, \infty) \rightarrow \mathbb{R}$ positive definite on $\mathbb{R}^d$ if the corresponding multivariate function $\Psi(\bm{x}) = \phi(\|\bm{x}\|_2)$, $\bm{x} \in \mathbb{R}^d$, is positive definite. That is, for any finite sequence of distinct vectors $\{\bm{x}_r\}_{r=1}^n$ in $\mathbb{R}^d$, the matrix $\bm{Z}$ defined as $\bm{Z}_{rs} = \Psi(\bm{x}_r - \bm{x}_s)$ is positive definite.
\end{definition}
The following theorem, which is a consequence of Schoenberg's theorem on the positive definiteness of completely monotone functions, provides the suitable tool for our purpose.

\begin{theorem}[Theorem 7.15 from \cite{wendland2004scat_data_approx}]
\label{thm:inverse_multiquadratics}
    The inverse multiquadrics $\phi(x) = \left(c^2 + x^2\right)^{-\beta}$ are positive definite functions on every $\mathbb{R}^d$, provided that $\beta > 0$ and $c > 0$.
\end{theorem}

\section{Supporting Results}
\label{sec:useful}

\begin{theorem}[Composite midpoint rule]
\label{thm:nonuniform_midpoint_2d}
Let \(a=x_0<x_1<\cdots <x_m=b\) and
\(c=y_0<y_1<\cdots <y_n=d\) be arbitrary partitions. Define
\(h_i=x_i-x_{i-1}\), \(k_j=y_j-y_{j-1}\),
\(\mu_i=(x_{i-1}+x_i)/2\), and \(\nu_j=(y_{j-1}+y_j)/2\).
Let \(h_{\max}=\max_{1\le i\le m}h_i\) and
\(k_{\max}=\max_{1\le j\le n}k_j\). Assume that \(f\), \(f_x\), \(f_y\),
and \(f_{xy}\) are integrable on \([a,b]\times[c,d]\), with the
absolute-continuity conditions needed for integration by parts on each
subrectangle. Suppose further that \(|\nabla f(x,y)|\le C_1\) and
\(|f_{xy}(x,y)|\le C_2\) on \([a,b]\times[c,d]\). Then
\[
\int_a^b\int_c^d f(x,y)\,dy\,dx
=
\sum_{i=1}^m\sum_{j=1}^n h_i k_j f(\mu_i,\nu_j)
+
E(f),
\]
where
\begin{align}
|E(f)|
&\le
\frac{C_1}{4}(b-a)(d-c)(h_{\max}+k_{\max})
\nonumber\\
&\quad+
\frac{C_2}{16}(b-a)(d-c)h_{\max}k_{\max}.
\label{eq:nonuniform_midpoint_error}
\end{align}
\end{theorem}

\begin{proof}
The proof is identical to the proof of Corollary~4.2 in
\cite{Grant2021numerical_integral}, except that the uniform cell widths
\(\Delta x\) and \(\Delta y\) are replaced by the nonuniform widths
\(h_i=x_i-x_{i-1}\) and \(k_j=y_j-y_{j-1}\). Applying the same local
midpoint-rule error identity on each rectangle
\([x_{i-1},x_i]\times[y_{j-1},y_j]\) and summing over all \(i,j\) gives
\begin{align*}
|E(f)|
&\le
\frac{C_1}{4}(d-c)\sum_{i=1}^m h_i^2
+
\frac{C_1}{4}(b-a)\sum_{j=1}^n k_j^2 \\
&\quad+
\frac{C_2}{16}
\left(\sum_{i=1}^m h_i^2\right)
\left(\sum_{j=1}^n k_j^2\right).
\end{align*}
Since \(\sum_i h_i=b-a\), \(\sum_j k_j=d-c\),
\(\sum_i h_i^2\le h_{\max}(b-a)\), and
\(\sum_j k_j^2\le k_{\max}(d-c)\), the stated bound follows.
\end{proof}

\begin{lemma}
\label{lem:trignometric_integral}
    Let $0<\delta_m<\pi/2$, then the following equality holds:
    \begin{equation}
    \label{eq:trignometric_integral}
        \int_{-\pi+\delta_m}^{-\delta_m} \int_{\delta_m}^{\pi-\delta_m} \frac{1}{2 - 2 \cos(\theta - \tau)} \; d\theta d\tau = -2 \ln \sin \delta_m.
    \end{equation}
\end{lemma}
\begin{proof}
The proof is a straightforward mathematical calculation. The LHS of \eqref{eq:trignometric_integral} is simplified as follows:
\begin{equation*}
\begin{aligned}
    &= \int_{-\pi+\delta_m}^{-\delta_m} \int_{\delta_m}^{\pi-\delta_m} \frac{1}{4 \sin^2 (\frac{\theta - \tau}{2})} d \theta d \tau \\
    &= \int_{-\pi+\delta_m}^{-\delta_m} \left( \frac{-1}{2} \cot (\frac{\theta -\tau}{2}) \right)_{\theta = \delta_m}^{\theta=\pi - \delta_m} d \tau \\
    &= \frac{1}{2} \left( \int_{-\pi+\delta_m}^{-\delta_m} \cot (\frac{\delta_m -\tau}{2}) d \tau - \int_{-\pi+\delta_m}^{-\delta_m} \tan(\frac{ \delta_m + \tau}{2}) d \tau \right).
\end{aligned}
\end{equation*}
We can calculate the integrals as follows:
\begin{equation*}
\begin{aligned}
    &= \frac{1}{2} \left( \left( -2\ln\sin(\frac{\delta_m - \tau}{2}) \right)_{-\pi+\delta_m}^{-\delta_m} \right. \\
    & \qquad\qquad \left. - \left( -2 \ln \cos(\frac{\tau + \delta_m}{2}) \right)_{-\pi+\delta_m}^{-\delta_m} \right) \\
    &= \frac{1}{2} \left( -2\ln\sin(\delta_m) + 2\ln\sin(\frac{\pi}{2}) + \right. \\ 
    & \qquad\qquad \left. 2 \ln \cos(0) - 2 \ln \cos(\delta_m - \frac{\pi}{2}) \right) \\
    &= -2 \ln \sin (\delta_m).
\end{aligned}
\end{equation*}
\end{proof}

\noindent\textit{Proof of Lemma~\ref{lem:L_F_to_u_2_norm_bound}.}

\begin{proof}
The Frobenius norm of the Loewner operator is given by
\begin{equation*}
    \norm{L(\bm{w})}_F^2 = \sum_{r=1}^M \sum_{s=1}^M \left| \frac{\bar{w}_r - w_s}{\bar{z}_r - z_s} \right|^2.
\end{equation*}
This summation can be approximated by a double integral of the function $f(\theta, \tau) = \left| \frac{G(e^{\bm{j} \tau}) - G(e^{\bm{j} \theta})}{e^{\bm{j} \tau} - e^{\bm{j} \theta}} \right|^2$,
defined over the region $R = \{ (\theta, \tau) \mid \theta \in [\delta_m, \pi - \delta_m], \; \tau \in [-\pi+\delta_m, -\delta_m] \}$.
Since $G$ is a rational function with no poles on the unit circle (recall that $G$ is stable), it is smooth. Denote by $C_r$ the upper bound on the absolute value of all of the $r$-th partial derivatives of $f$ over the compact region $R$. Since $f(\theta_r,\tau_s)\le
\tfrac{1}{h_{\min}^2}h_r h_s f(\theta_r,\tau_s)$, we have
\[
\norm{L(\bm{w})}_F^2
= \sum_{r=1}^M\sum_{s=1}^M f(\theta_r,\tau_s)
\le \frac{1}{h_{\min}^2}\sum_{r=1}^M\sum_{s=1}^M h_r h_s f(\theta_r,\tau_s).
\]
Applying Theorem~\ref{thm:nonuniform_midpoint_2d} to the weighted sum
(with $[a,b]=[\delta_m,\pi-\delta_m]$, $[c,d]=[-\pi+\delta_m,-\delta_m]$,
$h_{\max}=k_{\max}$), we obtain
\begin{equation}
\label{eq:Loewner_Frob_integ_approx}
\begin{aligned}
    &\norm{L(\bm{w})}_F^2 \\
    &\leq \frac{1}{h_{\min}^2} \int_{-\pi+\delta_m}^{-\delta_m} \int_{\delta_m}^{\pi-\delta_m}
    \left|\frac{G\left(e^{\bm{j} \tau}\right)-G\left(e^{\bm{j} \theta}\right)}
    {e^{\bm{j} \tau}-e^{\bm{j} \theta}}\right|^2 d \theta \, d \tau \\
    &\qquad\qquad
    + \frac{C_1(\pi - 2\delta_m)^2 h_{\max}}{2\, h_{\min}^2}
    + \frac{C_2(\pi - 2\delta_m)^2 h_{\max}^2}{16\, h_{\min}^2}.
\end{aligned}
\end{equation}
We aim to find an upper bound for the integral in \eqref{eq:Loewner_Frob_integ_approx}. manipulate the integrand as follows
\begin{align}
    &\left|\frac{G\left(e^{\bm{j} \tau}\right)-G\left(e^{\bm{j} \theta}\right)}{e^{\bm{j} \tau}-e^{\bm{j} \theta}}\right|^2 = \frac{\left| \sum_{k=0}^\infty g_k e^{\bm{j}k\tau} - \sum_{k=0}^\infty g_k e^{\bm{j}k\theta} \right|^2}{2 - e^{\bm{j} (\tau - \theta)} - e^{\bm{j} (\theta - \tau)}} \nonumber \\
    &=  \frac{\left| \sum_{k=0}^\infty g_k (e^{\bm{j} k\tau} - e^{\bm{j} k\theta}) \right|^2}{2 - 2 \cos(\theta - \tau)} \nonumber \\
    &= \frac{ \left(\sum_{p=0}^\infty g_p (e^{-\bm{j} p\tau} - e^{-\bm{j} p\theta})\right) \left(\sum_{q=0}^\infty g_q (e^{\bm{j} q\tau} - e^{\bm{j} q\theta})\right)}{2 - 2 \cos(\theta - \tau)} \nonumber \\
    &= \frac{ \sum_{p=0}^\infty \sum_{q=0}^\infty g_p g_q \left( e^{\bm{j}\tau(q-p)} + e^{\bm{j}\theta(q-p)} \right. }{2 - 2 \cos(\theta - \tau)} \cdots \nonumber \\
    &\hspace{10em} - \frac{ \left. e^{\bm{j}(q\tau-p\theta)} + e^{\bm{j}(q\theta-p\tau)} \right)}{2 - 2 \cos(\theta - \tau)}.
    \label{eq:exp_to_co_simp}
\end{align}
Each complex exponential term is bounded by 1, and thus, by the application of the triangle inequality, their summation is bounded by four. This results in the inequality
\begin{equation*}
    \left|\frac{G\left(e^{\bm{j} \tau}\right)-G\left(e^{\bm{j} \theta}\right)}{e^{\bm{j} \tau}-e^{\bm{j} \theta}}\right|^2 \leq
    \frac{ 4 \left| \sum_{p=0}^\infty \sum_{q=0}^\infty g_p g_q \right| }{2 - 2 \cos(\theta - \tau)}.
\end{equation*}
Breaking the double summation into two parts, we obtain
\begin{align}
    &\leq \frac{4 \left| \sum_{p=0}^\infty (g_p^2) + 2 \sum_{p=0}^\infty \sum_{s=1}^\infty (g_p g_{p+k}) \right|}{2 - 2 \cos(\theta - \tau)} \nonumber \\
    &\leq \frac{4 \left| \sum_{p=0}^\infty (g_p^2) + 2 \sum_{p=0}^\infty \sum_{k=1}^\infty (g_p^2 K_G \rho^{-k}) \right|}{2 - 2 \cos(\theta - \tau)} \nonumber \\
    &\leq \frac{4 \left( 1 + \frac{2K_G}{\rho - 1} \right) \sum_{p=0}^\infty (g_p^2)}{2 - 2 \cos(\theta - \tau)} = \frac{4 \left( 1 + \frac{2K_G}{\rho - 1} \right) \|\bm{g}\|_2^2}{2 - 2 \cos(\theta - \tau)}.
    \label{eq:integrand_upper_bound}
\end{align}
Because the integrand in \eqref{eq:Loewner_Frob_integ_approx} is upper bounded by \eqref{eq:integrand_upper_bound}, we find the following upper bound

\begin{align}
    &\norm{L(\bm{w})}_F^2 \leq \nonumber \\
    &\frac{4 \left( 1 + \frac{2K_G}{\rho - 1} \right) \|\bm{g}\|_2^2}{h_{\min}^2}
    \int_{-\pi+\delta_m}^{-\delta_m} \int_{\delta_m}^{\pi-\delta_m}
    \frac{1}{2 - 2 \cos(\theta - \tau)} \, d \theta \, d \tau \nonumber \\
    & + \frac{C_1(\pi - 2\delta_m)^2 h_{\max}}{2\, h_{\min}^2}
    + \frac{C_2(\pi - 2\delta_m)^2 h_{\max}^2}{16\, h_{\min}^2} \nonumber \\
    &= \frac{-8 \ln \sin \delta_m}{h_{\min}^2}
    \left( 1 + \frac{2K_G}{\rho - 1} \right) \|\bm{g}\|_2^2 \nonumber \\
    &\quad + \frac{C_1(\pi - 2\delta_m)^2 h_{\max}}{2\, h_{\min}^2}
    + \frac{C_2(\pi - 2\delta_m)^2 h_{\max}^2}{16\, h_{\min}^2}
    \label{eq:numerator_upper_bound}
\end{align}
in which we leveraged Lemma~\ref{lem:trignometric_integral} to take the integral. On the other hand, by Parseval's theorem, we have
\begin{equation*}
    \| \bm{g} \|_2^2 = \frac{1}{2 \pi} \int_{-\pi}^{\pi} | G(e^{\bm{j} \theta}) |^2 \; d\theta.
\end{equation*}
Since $\norm{\bm{w}}_2^2 = \sum_r |w_r|^2
\ge \tfrac{1}{h_{\max}}\sum_r h_r|w_r|^2$,
applying the composite midpoint rule
(1 dimensional version of Theorem~\ref{thm:nonuniform_midpoint_2d}) gives
\begin{align*}
\frac{1}{h_{\max}}\sum_r h_r|w_r|^2
&\ge \frac{1}{h_{\max}}
\left(\int_{\delta_m}^{\pi-\delta_m}|G(e^{\bm{j}\theta})|^2\,d\theta\right.\\
&\qquad\left.- \frac{D_2(\pi-2\delta_m)h_{\max}^2}{24}\right),
\end{align*}
where $D_2$ is an upper bound on the second derivative of $|G(e^{\bm{j}\theta})|^2$
over the unit circle. Using $D_0 = \sup_\theta |G(e^{\bm{j}\theta})|^2$ to bound
the truncated tails via
\[
\int_{\delta_m}^{\pi-\delta_m}|G(e^{\bm{j}\theta})|^2\,d\theta
\ge \pi\|\bm{g}\|_2^2 - 2D_0\delta_m,
\]
we obtain
\begin{equation}
\label{eq:h2norm_lower_bound}
    \norm{\bm{w}}_2^2 \;\ge\;
    \frac{\pi\|\bm{g}\|_2^2 - 2D_0\delta_m
    - D_2(\pi-2\delta_m)h_{\max}^2/24}{h_{\max}}.
\end{equation}

Combining \eqref{eq:numerator_upper_bound} and \eqref{eq:h2norm_lower_bound} results in
\begin{align}
&\frac{\norm{L(\bm{w})}_F^2}{\| \bm{w} \|_2^2} \leq \frac{h_{\max}}{h_{\min}^2}\cdot \nonumber \\
&\frac{\splitfrac{-8\ln\sin\delta_m\left(1+\frac{2K_G}{\rho-1}\right)\|\bm{g}\|_2^2 + \frac{C_1(\pi-2\delta_m)^2 h_{\max}}{2}}{ + \frac{C_2(\pi-2\delta_m)^2 h_{\max}^2}{16}}}{\pi\|\bm{g}\|_2^2 - 2D_0\delta_m - \frac{D_2(\pi-2\delta_m)h_{\max}^2}{24}}.\label{eq:fraction_bound}
\end{align}
Choose $h_{\max}$ and $\delta_m$ small enough (remember $h_{\max}$ can be decreased by choosing more frequency points $M$) such that for some $0 < \epsilon < 1$:
\begin{itemize}
    \item The denominator satisfies $\pi\|\bm{g}\|_2^2 - 2D_0\delta_m - \frac{D_2(\pi-2\delta_m)h_{\max}^2}{24} \geq (1-\epsilon)\pi\|\bm{g}\|_2^2$.
    \item The numerator satisfies $\frac{C_1(\pi-2\delta_m)^2 h_{\max}}{2} + \frac{C_2(\pi-2\delta_m)^2 h_{\max}^2}{16} \leq \epsilon\,(-8\ln\sin\delta_m)\left(1+\frac{2K_G}{\rho-1}\right)\|\bm{g}\|_2^2$.
\end{itemize}
Explicitly, the two conditions hold when $\delta_m$ and $h_{\max}$
are small enough:
\begin{itemize}
    \item \textbf{Condition on $\delta_m$:}
    \[
    \delta_m \leq \frac{\epsilon\pi\|\bm{g}\|_2^2}{4D_0},
    \]
    which ensures $2D_0\delta_m \leq \tfrac{\epsilon}{2}\pi\|\bm{g}\|_2^2$.
    \item \textbf{Conditions on $h_{\max}$:}
    \begin{align*}
    h_{\max} \leq \min\!\left(\vphantom{\frac{A}{B}}
    \sqrt{\frac{12\epsilon\pi\|\bm{g}\|_2^2}{D_2(\pi-2\delta_m)}},\right.\\
    \left.
    \frac{
    32\epsilon(-\ln\sin\delta_m)
    \left(1+\tfrac{2K_G}{\rho-1}\right)\|\bm{g}\|_2^2
    }
    {
    C_1(\pi-2\delta_m)^2
    +
    \sqrt{I_1+I_2}
    }
    \right),
    \end{align*}
    where $I_1 = C_1^2(\pi-2\delta_m)^4$ and $I_2 =
    8\epsilon C_2(\pi-2\delta_m)^2
    (-\ln\sin\delta_m)
    \left(1+\tfrac{2K_G}{\rho-1}\right)\|\bm{g}\|_2^2$.
    The first component ensures
    \[
    \frac{D_2(\pi-2\delta_m)h_{\max}^2}{24}
    \leq
    \tfrac{\epsilon}{2}\pi\|\bm{g}\|_2^2;
    \]
    the second ensures the numerator error terms are at most
    \[
    \epsilon(-8\ln\sin\delta_m)
    \left(1+\tfrac{2K_G}{\rho-1}\right)\|\bm{g}\|_2^2 .
    \]
\end{itemize}
Under both conditions, $\|\bm{g}\|_2^2$ cancels from
\eqref{eq:fraction_bound}:
\[
\frac{\norm{L(\bm{w})}_F^2}{\norm{\bm{w}}_2^2}
\leq \frac{-8h_{\max}\ln\sin\delta_m}{\pi h_{\min}^2}
\left(1+\frac{2K_G}{\rho-1}\right)\frac{1+\epsilon}{1-\epsilon}.
\]
This upper bound holds for any $0 < \epsilon < 1$. In particular, choosing $\epsilon = \frac{1}{17}$ establishes the bound stated in the Lemma.
\end{proof}

\noindent\textit{Proof of Lemma~\ref{lem:sum_z_bound_int}.}

\begin{proof}
Since $z_r = e^{\bm{j}\theta_r}$ and $\bar{z}_r = e^{-\bm{j}\theta_r}$,
setting $\tau_r = -\theta_r$ gives
$|\bar{z}_r - z_s|^2 = |e^{\bm{j}\tau_r} - e^{\bm{j}\theta_s}|^2
= 2 - 2\cos(\theta_s - \tau_r)$,
so the LHS of \eqref{eq:sum_1_overzsq_leq_lnsin_delta} equals
\[
\sum_{r=1}^M\sum_{s=1}^M \frac{1}{2-2\cos(\theta_s-\tau_r)}.
\]
Since $h_r \geq h_{\min}$ for all $r$, we have
$\sum_{r,s}\frac{1}{|\bar{z}_r-z_s|^2}
\leq \frac{1}{h_{\min}^2}\sum_{r,s} h_r h_s\,\frac{1}{|\bar{z}_r-z_s|^2}$.
The function $\frac{1}{2-2\cos(\theta-\tau)}$ is convex on $R$, so
for any partition the composite midpoint sum underestimates the integral:
\[
\sum_{r,s} h_r h_s\,\frac{1}{|\bar{z}_r-z_s|^2}
\leq \iint_R \frac{d\theta\,d\tau}{2-2\cos(\theta-\tau)}.
\]
Therefore
$\sum_{r,s}\frac{1}{|\bar{z}_r-z_s|^2}
\leq \frac{1}{h_{\min}^2}\iint_R \frac{d\theta\,d\tau}{2-2\cos(\theta-\tau)}$,
and the proof is complete by Lemma~\ref{lem:trignometric_integral}.
\end{proof}

\begin{lemma}
\label{lem:alg_connc_lower_bound}
    Let $L$ be the Laplacian of a weighted graph with $n$ nodes and denote the weight between nodes $r$ and $s$ with $w_{rs}$. Assume $w_{rs} \geq c$ for some $c > 0$. Denote the $r$-th eigenvalue of $L$ by $\lambda_r(L)$. The following bound holds:
    \begin{equation}
        \lambda_{n-1}(L) \geq cn.
    \end{equation}
\end{lemma}

\begin{proof}
    From the Rayleigh quotient \cite{mohar1991laplacian} we have
    \begin{equation*}
    \begin{aligned}
        \lambda_{n-1}(L) &= \min_{\|x\|_2 = 1, x^\top \bm{1}_n = 0} x^\top L x \\
        &= \min_{\|x\|_2 = 1, x^\top \bm{1}_n = 0} \sum_{r,s} w_{rs} (x_r - x_s)^2 \\
        &\geq c \min_{\|x\|_2 = 1, x^\top \bm{1}_n = 0} \sum_{r,s} (x_r - x_s)^2 = c \lambda_{n-1}(L_{comp})
    \end{aligned}
    \end{equation*}
    in which $L_{comp}$ is the Laplacian of a complete graph with $n$ nodes. The proof is complete by the fact that $\lambda_{n-1}(L_{comp}) = n$ \cite{mohar1991laplacian}.
\end{proof}

\begin{lemma}
\label{lem:pos_semi_def_kernel}
    Let $\bm{z} \in \mathbb{C}^M$ satisfy Assumption~\ref{ass:z_margin}. Then the matrix $\bm{X} \in \mathbb{C}^{M \times M}$ defined as $\bm{X}_{rs} = \frac{2}{|\bar{z}_r - z_s|^2}$ is positive semi-definite.
\end{lemma}
\begin{proof}
First, note that by Assumption~\ref{ass:z_margin}, there exists a frequency margin $\delta_m$ that ensures $\bm{X}$ is well-defined, as the denominator of $\frac{2}{|\bar{z}_r - z_s|^2}$ is nonzero. By applying Theorem~\ref{thm:inverse_multiquadratics}, the matrix $\bm{Y} \in \mathbb{C}^{M \times M}$, defined as  
\begin{equation*}
    \bm{Y}_{rs} = \frac{2}{c^2 + |\bar{z}_r - z_s|^2},
\end{equation*}  
is positive definite. On the other hand,  
\begin{equation*}
\begin{aligned}
    & \| \bm{X} - \bm{\bm{Y}} \|_2^2 \leq \| \bm{X} - \bm{Y} \|_F^2 \\
    &\leq \sum_{r,s=1}^M \left( \frac{2}{|\bar{z}_r - z_s|^2} - \frac{2}{c^2 + |\bar{z}_r - z_s|^2} \right)^2 \\
    &= \sum_{r,s=1}^M \frac{4c^4}{|\bar{z}_r - z_s|^4(|\bar{z}_r - z_s|^2 + c^2)^2} \\
    &\leq \frac{4Mc^4}{\delta_m^4(\delta_m^2 + c^2)^2}.
\end{aligned}
\end{equation*}
Thus, the following bound holds:
\begin{equation*}
    \| \bm{X} - \bm{Y} \|_2 \leq \frac{2c^2\sqrt{M}}{\delta_m^2(\delta_m^2 + c^2)},
\end{equation*}
where the RHS tends to zero as $c$ approaches zero. Therefore, since the eigenvalue as a function of the matrix is continuous, by choosing $c > 0$ sufficiently small, we can make the eigenvalues of $\bm{X}$ arbitrarily close to those of $\bm{Y}$, which completes the proof.
\end{proof}


\section {Adjoint of  the Loewner Operator and its Inverse}
\label{sec:adj_inv_loewner}

To compute the adjoint operator of $L$, define the Loewner matrices $\bm{E}_k$ and $\bm{F}_k$ for $k = 1, \cdots, M$, which capture the effect of unit perturbations in $\Re(w_k)$ and $\Im(w_k)$ on the Loewner matrix respectively, having nonzero entries only in the $k$-th row and column, as
\begin{equation}
\label{eq:def_E_k}
    (\bm{E}_k)_{rs} = \begin{cases}
        \frac{1}{\bar{z}_k - z_s} & r = k \text{ and } s \neq k \\
        \frac{-1}{\bar{z}_r - z_k} & r \neq k \text{ and } s = k \\
        0 & r=s=k \\
        0 & \text{otherwise}
    \end{cases}
\end{equation}
and
\begin{equation}
\label{eq:def_F_K}
    (\bm{F}_k)_{rs} = \begin{cases}
        \frac{-\bm{j}}{\bar{z}_k - z_s} & r = k \text{ and } s \neq k \\
        \frac{-\bm{j}}{\bar{z}_r - z_k} & r \neq k \text{ and } s = k \\
        \frac{-2\bm{j}}{\bar{z}_k - z_k} & r=s=k \\
        0 & \text{otherwise}.
    \end{cases}
\end{equation}
Clearly, $\mathcal{L}\left(\begin{bmatrix} \bm{e}_k \\ \bm{0}_M \end{bmatrix}\right) = \bm{E}_k$ and $\mathcal{L}\left(\begin{bmatrix} \bm{0}_M \\ \bm{e}_k \end{bmatrix}\right) = \bm{F}_k$, and thus, $\{\bm{E}_k, \bm{F}_k\}_{k=1}^M$ form a basis for the space of all Loewner matrices. Therefore, the \emph{adjoint Loewner operator} $\mathcal{L}^*:\mathbb{L}(M) \to \mathbb{C}^M$ is
\begin{equation}
\label{eq:loewner_adjoint}
    \mathcal{L}^*(\bm{X}) = \begin{bmatrix}
        \langle E_1, \bm{X} \rangle \\
        \vdots \\
        \langle E_M, \bm{X} \rangle \\
        \langle F_1, \bm{X} \rangle \\
        \vdots \\
        \langle F_M, \bm{X} \rangle \\
    \end{bmatrix}
\end{equation}
for all $\bm{X} \in \mathbb{L}(M)$.  

To find the inverse of the adjoint of the Loewner operator, we follow the definition of the inverse operator. $\bm{X} = \LL^{-*}(\begin{bmatrix} \bm{a} \\ \bm{b} \end{bmatrix})$ iff $L^*(\bm{X}) = \begin{bmatrix} \bm{a} \\ \bm{b} \end{bmatrix}$. Thus,
\begin{equation}
\label{eq:L^-*}
        \LL^{-*}(\begin{bmatrix} \bm{a} \\ \bm{b} \end{bmatrix}) = \bm{X} \text{ iff } a_k = \langle \bm{E}_k, \bm{X}\rangle \text{ and } b_k = \langle \bm{F}_k, \bm{X}\rangle.
\end{equation}
Because $\{\bm{E}_k, \bm{F}_k\}_{k=1}^M$ is a basis for $\mathbb{L}(M)$, there exists unique $\bm{c}, \bm{d} \in \mathbb{R}^M$ such that $\bm{X} = \sum_{k=1}^M c_k \bm{E}_k + \sum_{k=1}^M d_k \bm{F}_k$. Thus, taking the inner products $\langle \bm{E}_r, \bm{X} \rangle$ and $\langle \bm{F}_r, \bm{X} \rangle$ results in the following system of linear equations:
\begin{equation}
\label{eq:E_inn_F_inn_cd_ab}
    \begin{bmatrix}
        \langle \bm{E}_r, \bm{E}_k \rangle & \langle \bm{E}_r, \bm{F}_k \rangle \\
        \langle \bm{F}_r, \bm{E}_k \rangle & \langle \bm{F}_r, \bm{F}_k \rangle
    \end{bmatrix} \begin{bmatrix}
        \bm{c} \\ \bm{d}
    \end{bmatrix} = \begin{bmatrix}
        \bm{a} \\ \bm{b}
    \end{bmatrix}.
\end{equation}
Due to the structure of $\bm{E}_k$ and $\bm{F}_k$ described in \eqref{eq:def_E_k} and \eqref{eq:def_F_K}, the off-diagonal block matrices in the LHS of \eqref{eq:E_inn_F_inn_cd_ab} are zero. Thus, the \emph{inverse adjoint Loewner operator} is
\begin{equation}
\label{eq:loewner_inv_conj}
\begin{aligned}
    \LL^{-*}(\begin{bmatrix}
        \bm{a} \\ \bm{b}
    \end{bmatrix}) = \sum_{k=1}^M \left( \bm{e}_k^\top \Ein^{\dagger} \bm{a} \bm{E}_k \right) + \sum_{k=1}^M \left( \bm{e}_k^\top \Fin^{-1} \bm{b} \bm{F}_k \right).
\end{aligned}
\end{equation}
in which $\Ein, \Fin \in \Cxx{M}{M}$ are such that $ (\Ein)_{rk} = \langle \bm{E}_r, \bm{E}_k \rangle $ and $ (\Fin)_{rk} = \langle \bm{F}_r, \bm{F}_k \rangle $. Similar to the Loewner matrix, the matrix $ E_{\text{in}} $ is not invertible, and its kernel has a dimension of one with $\mathbf{1}_M$ as its basis. Using the pseudo-inverse of this matrix to find $\bm{c}$ yields a solution corresponding to the condition that $\sum c_k = 0$ following the construction of the invertible Loewner operator in Section~\ref{sec:loewner_operator}. From \eqref{eq:def_E_k} and \eqref{eq:def_F_K}, we have the followings:
\begin{equation}
\label{eq:E_in}
    (\Ein)_{rs} = \begin{cases}
        \frac{-2}{| \bar{z}_r - z_s |^2} & r \neq s \\
        \sum_{k \neq r} \frac{2}{| \bar{z}_r - z_k |^2} & r = s
    \end{cases}
\end{equation}
and
\begin{equation}
\label{eq:F_in}
    (\Fin)_{rs} = \begin{cases}
        \frac{2}{| \bar{z}_r - z_s |^2} & r \neq s \\
        \sum_{k \neq r} \frac{2}{| \bar{z}_r - z_k |^2} + \frac{4}{| \bar{z}_r - z_r |^2} & r = s.
    \end{cases}
\end{equation}

\section*{Acknowledgment}
The authors thank Reza Vafaee for helpful discussions.


\bibliographystyle{IEEEtran} 
\bibliography{references_MS}

@INPROCEEDINGS{tsiamis2024finitesamplefrequencydomain,
  author={Tsiamis, Anastasios and Abdalmoaty, Mohamed and Smith, Roy S. and Lygeros, John},
  booktitle={2024 IEEE 63rd Conference on Decision and Control (CDC)}, 
  title={Finite Sample Frequency Domain Identification}, 
  year={2024},
  volume={},
  number={},
  pages={3025-3030},
  doi={10.1109/CDC56724.2024.10885823}
}

@article{mayo2007framework_solution_generalized_realization,
title = {A framework for the solution of the generalized realization problem},
journal = {Linear Algebra and its Applications},
volume = {425},
number = {2},
pages = {634-662},
year = {2007},
note = {Special Issue in honor of Paul Fuhrmann},
issn = {0024-3795},
doi = {https://doi.org/10.1016/j.laa.2007.03.008},
author = {A.J. Mayo and A.C. Antoulas}
}

@ARTICLE{bhaskar2013atomic_norm_denoising_line_spectral_estimation,
  author={Bhaskar, Badri Narayan and Tang, Gongguo and Recht, Benjamin},
  journal={IEEE Transactions on Signal Processing}, 
  title={Atomic Norm Denoising With Applications to Line Spectral Estimation}, 
  year={2013},
  volume={61},
  number={23},
  pages={5987-5999},
  doi={10.1109/TSP.2013.2273443}}

@article{cai2015robust_recovery_exp_signals_gaussian_hankel,
title = {Robust recovery of complex exponential signals from random Gaussian projections via low rank Hankel matrix reconstruction},
journal = {Applied and Computational Harmonic Analysis},
volume = {41},
number = {2},
pages = {470-490},
year = {2016},
issn = {1063-5203},
doi = {https://doi.org/10.1016/j.acha.2016.02.003},
url = {https://www.sciencedirect.com/science/article/pii/S1063520316000117},
author = {Jian-Feng Cai and Xiaobo Qu and Weiyu Xu and Gui-Bo Ye}
}

@ARTICLE{fazel2022finite_sample_system_identification_nuclear_norm,
  author={Sun, Yue and Oymak, Samet and Fazel, Maryam},
  journal={IEEE Open Journal of Control Systems}, 
  title={Finite Sample Identification of Low-Order LTI Systems via Nuclear Norm Regularization}, 
  year={2022},
  volume={1},
  number={},
  pages={237-254},
  doi={10.1109/OJCSYS.2022.3200015}
}

@article{Grant2021numerical_integral,
place={Minneapolis, MN},
title={Elementary Numerical Methods for Double Integrals}, volume={4},
url={https://pubs.lib.umn.edu/index.php/mjum/article/view/4157},
number={1},
journal={Minnesota Journal of Undergraduate Mathematics},
author={Grant, Cameron Jason and Talvila, Erik},
year={2021},
month={May}
}

@book{hastie2009esl,
  title={The elements of statistical learning: data mining, inference, and prediction},
  author={Hastie, Trevor and Tibshirani, Robert and Friedman, Jerome H and Friedman, Jerome H},
  volume={2},
  year={2009},
  publisher={Springer}
}

@ARTICLE{chen1995Hinf_interpol_sysid,
  author={Jie Chen and Nett, C.N. and Fan, M.K.H.},
  journal={IEEE Transactions on Automatic Control}, 
  title={Worst case system identification in H/sub /spl infin//: validation of a priori information, essentially optimal algorithms, and error bounds},
  year={1995},
  volume={40},
  number={7},
  pages={1260-1265},
  doi={10.1109/9.400481}
}

@book{keesman2011sysid,
  title={System identification: an introduction},
  author={Keesman, Karel J},
  year={2011},
  publisher={Springer Science \& Business Media}
}

@article{chiuso2019sysid_learning,
  title={System identification: A machine learning perspective},
  author={Chiuso, Alessandro and Pillonetto, Gianluigi},
  journal={Annual Review of Control, Robotics, and Autonomous Systems},
  volume={2},
  number={1},
  pages={281--304},
  year={2019},
  publisher={Annual Reviews}
}

@inproceedings{simchowitz2018sysid_lti_mixing,
  title={Learning without mixing: Towards a sharp analysis of linear system identification},
  author={Simchowitz, Max and Mania, Horia and Tu, Stephen and Jordan, Michael I and Recht, Benjamin},
  booktitle={Conference On Learning Theory},
  pages={439--473},
  year={2018},
  organization={PMLR}
}

@article{tu2024learning_many_trajectories,
  title={Learning from many trajectories},
  author={Tu, Stephen and Frostig, Roy and Soltanolkotabi, Mahdi},
  journal={Journal of Machine Learning Research},
  volume={25},
  number={216},
  pages={1--109},
  year={2024}
}

@article{tropp2012random_matrices,
  title={User-friendly tail bounds for sums of random matrices},
  author={Tropp, Joel A},
  journal={Foundations of computational mathematics},
  volume={12},
  pages={389--434},
  year={2012},
  publisher={Springer}
}

@article{tsiamis2023statistical_learning_control,
  title={Statistical learning theory for control: A finite-sample perspective},
  author={Tsiamis, Anastasios and Ziemann, Ingvar and Matni, Nikolai and Pappas, George J},
  journal={IEEE Control Systems Magazine},
  volume={43},
  number={6},
  pages={67--97},
  year={2023},
  publisher={IEEE}
}

@article{fazel2013hankel__rank_sysid,
  title={Hankel matrix rank minimization with applications to system identification and realization},
  author={Fazel, Maryam and Pong, Ting Kei and Sun, Defeng and Tseng, Paul},
  journal={SIAM Journal on Matrix Analysis and Applications},
  volume={34},
  number={3},
  pages={946--977},
  year={2013},
  publisher={SIAM}
}

@inproceedings{fazel2001rank_minimization_min_order,
  title={A rank minimization heuristic with application to minimum order system approximation},
  author={Fazel, Maryam and Hindi, Haitham and Boyd, Stephen P},
  booktitle={Proceedings of the 2001 American control conference.(Cat. No. 01CH37148)},
  volume={6},
  pages={4734--4739},
  year={2001},
  organization={IEEE}
}

@article{khosravi2020lowcomplexity_sparse_estimation,
  title={Low-complexity identification by sparse hyperparameter estimation},
  author={Khosravi, Mohammad and Yin, Mingzhou and Iannelli, Andrea and Parsi, Anilkumar and Smith, Roy S},
  journal={IFAC-PapersOnLine},
  volume={53},
  number={2},
  pages={412--417},
  year={2020},
  publisher={Elsevier}
}

@ARTICLE{smith2014frequency_nuc_norm,
  author={Smith, Roy S.},
  journal={IEEE Transactions on Automatic Control}, 
  title={Frequency Domain Subspace Identification Using Nuclear Norm Minimization and Hankel Matrix Realizations}, 
  year={2014},
  volume={59},
  number={11},
  pages={2886-2896},
  doi={10.1109/TAC.2014.2351731}
}

@article{pillonetto2016regularized_sysid,
  title={Regularized linear system identification using atomic, nuclear and kernel-based norms: The role of the stability constraint},
  author={Pillonetto, Gianluigi and Chen, Tianshi and Chiuso, Alessandro and De Nicolao, Giuseppe and Ljung, Lennart},
  journal={Automatica},
  volume={69},
  pages={137--149},
  year={2016},
  publisher={Elsevier}
}

@article{mckelvey2002frequency,
  title={Frequency domain identification methods},
  author={McKelvey, Tomas},
  journal={Circuits, Systems and Signal Processing},
  volume={21},
  number={1},
  pages={39--55},
  year={2002},
  publisher={Springer}
}

@article{mckelvey2000frequency_low_order,
  title={Frequency domain identification},
  author={McKelvey, Tomas},
  journal={IFAC Proceedings Volumes},
  volume={33},
  number={15},
  pages={7--18},
  year={2000},
  publisher={Elsevier}
}

@INPROCEEDINGS{jedra2019sample_complexity,
  author={Jedra, Yassir and Proutiere, Alexandre},
  booktitle={2019 IEEE 58th Conference on Decision and Control (CDC)}, 
  title={Sample Complexity Lower Bounds for Linear System Identification}, 
  year={2019},
  volume={},
  number={},
  pages={2676-2681},
  doi={10.1109/CDC40024.2019.9029303}
}

@ARTICLE{fattahi2021sample_complexity,
  author={Fattahi, Salar and Sojoudi, Somayeh},
  journal={IEEE Transactions on Control of Network Systems}, 
  title={Sample Complexity of Block-Sparse System Identification Problem}, 
  year={2021},
  volume={8},
  number={4},
  pages={1905-1917},
  doi={10.1109/TCNS.2021.3089141}
}

@ARTICLE{Chatzikiriakos2024sample_complexity,
  author={Chatzikiriakos, Nicolas and Iannelli, Andrea},
  journal={IEEE Control Systems Letters}, 
  title={Sample Complexity Bounds for Linear System Identification From a Finite Set},
  year={2024},
  volume={8},
  number={},
  pages={2751-2756},
  doi={10.1109/LCSYS.2024.3514995}
}

@inproceedings{sarkar2019sample_complexity,
  title={Near optimal finite time identification of arbitrary linear dynamical systems},
  author={Sarkar, Tuhin and Rakhlin, Alexander},
  booktitle={International Conference on Machine Learning},
  pages={5610--5618},
  year={2019},
  organization={PMLR}
}

@article{oymak2021sample_complexity_ho_kalman,
  title={Revisiting ho--kalman-based system identification: Robustness and finite-sample analysis},
  author={Oymak, Samet and Ozay, Necmiye},
  journal={IEEE Transactions on Automatic Control},
  volume={67},
  number={4},
  pages={1914--1928},
  year={2021},
  publisher={IEEE}
}

@article{zheng2020sample_complexity,
  title={Non-asymptotic identification of linear dynamical systems using multiple trajectories},
  author={Zheng, Yang and Li, Na},
  journal={IEEE Control Systems Letters},
  volume={5},
  number={5},
  pages={1693--1698},
  year={2020},
  publisher={IEEE}
}

@InProceedings{ziemann2022sample_complexity,
  title = 	 {Single Trajectory Nonparametric Learning of Nonlinear Dynamics},
  author =       {Ziemann, Ingvar M and Sandberg, Henrik and Matni, Nikolai},
  booktitle = 	 {Proceedings of Thirty Fifth Conference on Learning Theory},
  pages = 	 {3333--3364},
  year = 	 {2022},
  editor = 	 {Loh, Po-Ling and Raginsky, Maxim},
  volume = 	 {178},
  series = 	 {Proceedings of Machine Learning Research},
  month = 	 {02--05 Jul},
  publisher =    {PMLR},
  url = 	 {https://proceedings.mlr.press/v178/ziemann22a.html}
}

@incollection{karachalios2021loewner_framework_sysid_mor,
  title={The Loewner framework for system identification and reduction},
  author={Karachalios, Dimitrios and Gosea, Ion Victor and Antoulas, Athanasios C},
  booktitle={Model Order Reduction: Volume I: System-and Data-Driven Methods and Algorithms},
  pages={181--228},
  year={2021},
  publisher={de Gruyter}
}

@article{antoulas2017tutorial_loewner,
  title={A tutorial introduction to the Loewner framework for model reduction},
  author={Antoulas, Athanasios C and Lefteriu, Sanda and Ionita, A Cosmin and Benner, P and Cohen, A},
  journal={Model Reduction and Approximation: Theory and Algorithms},
  volume={15},
  pages={335},
  year={2017},
  publisher={SIAM Philadelphia}
}

@inproceedings{singh2020loewner_convex_low_order,
  title={A loewner matrix based convex optimization approach to finding low rank mixed time/frequency domain interpolants},
  author={Singh, Rajiv and Sznaier, Mario},
  booktitle={2020 American Control Conference (ACC)},
  pages={5169--5174},
  year={2020},
  organization={IEEE}
}

@article{honarpisheh2024loewner,
  title={Identification of Low Order Systems in a Loewner Framework},
  author={Honarpisheh, Arya and Singh, Rajiv and Miller, Jared and Sznaier, Mario},
  journal={IFAC-PapersOnLine},
  volume={58},
  number={15},
  pages={199--204},
  year={2024},
  publisher={Elsevier}
}

@book{wendland2004scat_data_approx,
  title={Scattered data approximation},
  author={Wendland, Holger},
  volume={17},
  year={2004},
  publisher={Cambridge university press}
}

@article{sznaier2014fast_nnm,
  title={Fast structured nuclear norm minimization with applications to set membership systems identification},
  author={Sznaier, Mario and Ayazoglu, Mustafa and Inanc, Tamer},
  journal={IEEE Transactions on Automatic Control},
  volume={59},
  number={10},
  pages={2837--2842},
  year={2014},
  publisher={IEEE}
}

@article{yilmaz2017randomized_pars_iden,
  title={A randomized algorithm for parsimonious model identification},
  author={Y{\i}lmaz, Burak and Bekiroglu, Korkut and Lagoa, Constantino and Sznaier, Mario},
  journal={IEEE Transactions on Automatic Control},
  volume={63},
  number={2},
  pages={532--539},
  year={2017},
  publisher={IEEE}
}

@article{parrilo1998mixed_robust_iden,
  title={Mixed time/frequency-domain based robust identification},
  author={Parrilo, Pablo A and Sznaier, Mario and Pena, RS S{\'a}nchez and Inanc, Tamer},
  journal={Automatica},
  volume={34},
  number={11},
  pages={1375--1389},
  year={1998},
  publisher={Elsevier}
}

@article{watson1992characterization_subdifferentil_matrix_norm,
  title={Characterization of the subdifferential of some matrix norms},
  author={Watson, G Alistair},
  journal={Linear Algebra Appl},
  volume={170},
  number={1},
  pages={33--45},
  year={1992}
}

@article{candes2012exact_matrix_comp_convex,
  title={Exact matrix completion via convex optimization},
  author={Candes, Emmanuel and Recht, Benjamin},
  journal={Communications of the ACM},
  volume={55},
  number={6},
  pages={111--119},
  year={2012},
  publisher={ACM New York, NY, USA}
}

@article{recht2010guaranteed,
  title={Guaranteed minimum-rank solutions of linear matrix equations via nuclear norm minimization},
  author={Recht, Benjamin and Fazel, Maryam and Parrilo, Pablo A},
  journal={SIAM review},
  volume={52},
  number={3},
  pages={471--501},
  year={2010},
  publisher={SIAM}
}

@article{sattar2025finite_sample_part_bil,
  title={Finite sample identification of partially observed bilinear dynamical systems},
  author={Sattar, Yahya and Jedra, Yassir and Fazel, Maryam and Dean, Sarah},
  journal={arXiv preprint arXiv:2501.07652},
  year={2025}
}

@INPROCEEDINGS{mashhadireza2022detection,
  title           = "Detection of anomalous behavior of a scaled laboratory
                     structure based on stochastic subspace identification
                     method",
  booktitle       = "2022 10th {RSI} International Conference on Robotics and
                     Mechatronics ({ICRoM})",
  author          = "Mashhadireza, Ali and Dehnavi, Narges Dehghani and
                     Sadighi, Ali and Bitaraf, Maryam",
  publisher       = "IEEE",
  month           =  nov,
  year            =  2022,
  conference      = "2022 10th RSI International Conference on Robotics and
                     Mechatronics (ICRoM)",
  location        = "Tehran, Iran, Islamic Republic of"
}

@ARTICLE{Mashhadireza2025novelty,
  title     = "Novelty detection in laboratory structure using stochastic
               subspace identification and a novel non-destructive experimental
               method for damping changes",
  author    = "Mashhadireza, Ali and Sadighi, Ali and Bitaraf, Maryam and
               Dehghani Dehnavi, Narges",
  journal   = "Struct. Eng. Int.",
  publisher = "Informa UK Limited",
  volume    =  35,
  number    =  4,
  pages     = "604--616",
  month     =  oct,
  year      =  2025,
  language  = "en"
}

@inproceedings{lee2022improved_rates_partial_obs,
  title={Improved rates for prediction and identification of partially observed linear dynamical systems},
  author={Lee, Holden},
  booktitle={International Conference on Algorithmic Learning Theory},
  pages={668--698},
  year={2022},
  organization={PMLR}
}

@article{mohar1991laplacian,
  title={The Laplacian spectrum of graphs},
  author={Mohar, Bojan and Alavi, Y and Chartrand, G and Oellermann, Ortrud},
  journal={Graph theory, combinatorics, and applications},
  volume={2},
  number={871-898},
  pages={12},
  year={1991},
  publisher={Wiley}
}

@ARTICLE{mckelvey1996subspace_freq,
  author={McKelvey, T. and Akcay, H. and Ljung, L.},
  journal={IEEE Transactions on Automatic Control}, 
  title={Subspace-based multivariable system identification from frequency response data}, 
  year={1996},
  volume={41},
  number={7},
  pages={960-979},
  doi={10.1109/9.508900}}

@article{khojaste2025quantile,
  author  = {Khojaste, Arash and Pritchard, Geoffrey and Zakeri, Golbon},
  title   = {Quantile Fourier Regressions for Decision Making Under Uncertainty},
  journal = {Annals of Operations Research},
  volume  = {355},
  pages   = {2843--2858},
  year    = {2025},
  doi     = {10.1007/s10479-024-06306-9},
  url     = {https://doi.org/10.1007/s10479-024-06306-9}
}

@InProceedings{kanakeri2025outlier,
  title = 	 {Outlier-Robust Linear System Identification Under Heavy-Tailed Noise},
  author =       {Kanakeri, Vinay and Mitra, Aritra},
  booktitle = 	 {Proceedings of the 7th Annual Learning for Dynamics \&amp; Control Conference},
  pages = 	 {540--551},
  year = 	 {2025},
  editor = 	 {Ozay, Necmiye and Balzano, Laura and Panagou, Dimitra and Abate, Alessandro},
  volume = 	 {283},
  series = 	 {Proceedings of Machine Learning Research},
  month = 	 {04--06 Jun},
  publisher =    {PMLR},
  url = 	 {https://proceedings.mlr.press/v283/kanakeri25a.html}
}

@article{shen2023computationally,
author = {Yinan Shen and Jingyang Li and Jian-Feng Cai and Dong Xia},
title = {{Computationally efficient and statistically optimal robust high-dimensional linear regression}},
volume = {53},
journal = {The Annals of Statistics},
number = {1},
publisher = {Institute of Mathematical Statistics},
pages = {374 -- 399},
year = {2025},
doi = {10.1214/24-AOS2473},
URL = {https://doi.org/10.1214/24-AOS2473}
}

@book{vanoverschee1996subspace,
  author    = {Van Overschee, Peter and De Moor, Bart},
  title     = {Subspace Identification for Linear Systems: Theory, Implementation, Applications},
  publisher = {Kluwer Academic Publishers},
  year      = {1996},
  address   = {Boston, MA},
  isbn      = {978-0-7923-9717-5}
}

@article{Antoulas:1986,
  author  = {Antoulas, A. C. and Anderson, B. D. O.},
  title   = {On the scalar rational interpolation problem},
  journal = {IMA Journal of Mathematical Control and Information},
  volume  = {3},
  pages   = {61--88},
  year    = {1986}
}

@article{Mayo:2007,
  author  = {Mayo, A. J. and Antoulas, A. C.},
  title   = {A framework for the solution of the generalized realization problem},
  journal = {Linear Algebra and its Applications},
  volume  = {425},
  pages   = {634--662},
  year    = {2007}
}

@book{Antoulas:2010,
  author    = {Antoulas, A. C.},
  title     = {Approximation of Large-Scale Dynamical Systems},
  publisher = {SIAM},
  year      = {2005}
}

@article{Ionita:2013,
  author  = {Ionita, A. C. and Antoulas, A. C.},
  title   = {Data-driven parametrized model reduction in the {L}oewner framework},
  journal = {SIAM Journal on Scientific Computing},
  volume  = {36},
  pages   = {A984--A1007},
  year    = {2014}
}

@book{garnett1981bounded,
  author    = {Garnett, John B.},
  title     = {Bounded Analytic Functions},
  series    = {Pure and Applied Mathematics},
  volume    = {96},
  publisher = {Academic Press},
  address   = {New York},
  year      = {1981},
 }

@book{pintelon2012system,
  title={System identification: a frequency domain approach},
  author={Pintelon, Rik and Schoukens, Johan},
  year={2012},
  publisher={John Wiley \& Sons}
}

@misc{fditools_github,
  title        = {{FdiTools}: Frequency Domain System Identification {MATLAB} Toolbox},
  author       = {{HoriFujimotoLab}},
  year         = {2024},
  howpublished = {\url{https://github.com/HoriFujimotoLab/FdiTools}},
  note         = {Accessed: 2026-06-15}
}

@misc{matlab_sysid_toolbox,
  author       = {{The MathWorks, Inc.}},
  title        = {System Identification Toolbox},
  year         = {2024},
  howpublished = {\url{https://www.mathworks.com/products/sysid.html}},
  note         = {Natick, Massachusetts, United States}
}

\end{document}